\documentclass[prd,preprint,eqsecnum,nofootinbib,amsmath,amssymb,
               tightenlines,dvips]{revtex4}
\usepackage{graphicx}
\usepackage{bm}

\begin {document}


\def\Mrowczynski{Mr\'owczy\'nski}
\def\Qs{Q_{\rm s}}

\def\alphas{\alpha_{\rm s}}
\def\hard{{\rm hard}}
\def\soft{{\rm soft}}
\def\D{{\bm D}}
\def\k{{\bm k}}
\def\half{{\textstyle{\frac12}}}
\def\fourth{{\textstyle{\frac14}}}
\def\e{{\bm e}}
\def\Amat{{\cal A}}

\def\p{{\bm p}}

\def\x{{\bm x}}
\def\y{{\bm y}}
\def\v{{\bm v}}
\def\E{{\bm E}}
\def\B{{\bm B}}
\def\A{{\bm A}}

\def\grad{{\bm\nabla}}

\def\tr{\operatorname{tr}}

\def\eps{\epsilon}



\title
    {
    The Abelianization of QCD Plasma Instabilities
    }

\author {Peter Arnold}
\author {Jonathan Lenaghan}
\affiliation
    {%
    Department of Physics,
    University of Virginia,
    P.O. Box 400714
    Charlottesville, VA 22901-4714
    }%

\date {August 2004}

\begin {abstract}%
    {%
       QCD plasma instabilities appear to play an important role in
       the equilibration of quark-gluon plasmas in heavy-ion collisions in
       the theoretical limit of weak coupling (i.e.\ asymptotically high
       energy).
       It is important to understand what non-linear physics
       eventually stops the exponential growth of unstable modes.
       It is already known that the initial growth of plasma instabilities
       in QCD closely parallels that in QED.  However, once the unstable
       modes of the gauge-fields grow large enough for non-Abelian
       interactions between them to become important, one might guess that
       the dynamics of QCD plasma instabilities and QED plasma instabilities
       become very different.
       In this paper, we give suggestive arguments that non-Abelian
       self-interactions between the unstable modes are ineffective at
       stopping instability growth, and that the growing non-Abelian
       gauge fields become approximately Abelian after a certain
       stage in their growth.
       This in turn suggests that understanding the development of
       QCD plasma instabilities in the non-linear regime may have
       close parallels to similar processes in traditional plasma physics.
       We conjecture that the physics of collisionless plasma instabilities
       in SU(2) and SU(3) gauge theory becomes equivalent, respectively,
       to (i) traditional plasma physics, which is  U(1) gauge theory,
       and (ii) plasma physics of
       U(1)$\times$U(1) gauge theory.
    }%
\end {abstract}

\maketitle
\thispagestyle {empty}


\section {Introduction}
\label{sec:intro}

At high enough energy, heavy ion collisions are presumed to create a
quark-gluon plasma in local equilibrium.  The physics associated with the
creation and equilibration of this plasma is rather complicated, involving
many different physical processes playing important roles at different stages
of the collision.  It is theoretically useful to try to understand the
collision in the formal limit of arbitrarily high energies, for which the
running coupling constant $\alphas$ is arbitrarily small due to
asymptotic freedom.  In this limit, it is believed that very early times are
described by the saturation scenario
\cite{gribov,blaizot,qiu,larry,krasnitz1,krasnitz2},
in which
the non-equilibrium plasma starts out at early times $\tau_0 \sim 1/\Qs$ as a
gas of gluons with
(i) non-perturbatively large phase-space density $f \sim 1/\alphas$ and
(ii) momenta of order a scale $\Qs$ known as the saturation
scale.
In the weak coupling limit $\alphas(\Qs) \ll 1$, Baier, Mueller,
Schiff and Son \cite{BMSS} attempted to find the parametric dependence of the
equilibration time $\tau_{\rm eq}$ on $\alphas$ and found $\tau_{\rm eq} \sim
\alpha^{-13/5} \Qs^{-1}$.  Their analysis is known as the bottom-up
thermalization scenario, and the result arises from a complicated interplay of
(i) the one-dimensional expansion of the plasma between the large,
Lorentz-contracted, nuclear pancakes after the collision
and (ii) a variety of individual collisional processes which relax the
plasma toward equilibrium.  However, collective processes are often more
important in plasmas than individual collisions.  In particular, \Mrowczynski\
and others
\cite{mrow0,mrow1,mrow2,mrow3,mrow&thoma,randrup&mrow,heinz_conf,
selikhov1,selikhov2,selikhov3,pavlenko}
have long suggested that plasma
instabilities might play an important role in the equilibration of quark-gluon
plasmas.  Arnold, Lenaghan, and Moore \cite{alm} showed that this is indeed
the case for the bottom-up analysis.  That means that the bottom-up
scenario of Baier {\it et al.}\/ needs to be completely reanalyzed.
The theory of quark-gluon plasma equilibration in heavy-ion
collisions is currently in the embarrassing state of not even knowing
how the equilibration time depends on $\alphas$ in the weak coupling
limit --- that is, not even the power $n$ is known in the parametric
relation
\begin {equation}
   \tau_{\rm eq} \sim \alpha^{-n} \Qs^{-1} .
\label {eq:taueq}
\end {equation}

For ultra-relativistic, homogeneous, parity-invariant,
collisionless plasmas, a long-wavelength magnetic instability known
as the Weibel or filamentary
instability generically arises whenever the momentum-distribution of charged
particles in the plasma is anisotropic.
A proof (and a more precise statement) may be found in Ref.\ \cite{alm}.
Reviews in the quark-gluon plasma literature
of the physical origin of the instability may be found in
Refs.\ \cite{mrow3,alm}.
The actual quark-gluon plasma in heavy-ion collisions is not collisionless,
nor precisely homogeneous, but Ref.\ \cite{alm} showed that the
instabilities that plague the original bottom-up scenario are associated with
small enough distance and time scales that collisions and inhomogeneity
can be ignored.%
\footnote{
   Whether this will continue to be true in some future bottom-up scenario
   that incorporates plasma instabilities from the beginning remains to
   be seen.
}
The analysis of such instabilities in the quark-gluon plasma literature
has generally been restricted to linearized analysis, which treats the
amplitude of the unstable long-wavelength magnetic field as perturbatively
small.  This is adequate for calculating the linear growth rate
of the instability.  In order to understand how instabilities affect
equilibration, however, we need to know just how big
the unstable long-wavelength magnetic field grows and what happens
next.  In particular, does the presence of this non-perturbatively large,
long-wavelength field cause rapid isotropization and equilibration of
the system?  And what (parametrically) are the time scales involved?
These questions cannot be answered by a linear analysis of the instability.
In this paper, we will modestly focus on the very first question: How
big does the instability grow?
In answering this question, we will also learn important lessons
concerning the qualitative nature of what happens to the unstable
modes of the theory once they have grown large.

To investigate collective effects, it is both standard and
convenient to describe the
non-equilibrium quark-gluon plasma using kinetic theory in the form
of Vlasov equations.
We split the system into two parts: short wavelength (``hard'')
momentum excitations which are
described by a Boltzmann equation, and long-wavelength (``soft'')
modes which are
described by classical Maxwell equations.
Hard excitations are treated as a collection
of particles with phase-space density $f(\p,\x,t)$.
For an Abelian theory, the corresponding (collisionless)
Boltzmann equation would be
\begin {equation}
   \left[\partial_t + \v\cdot\grad_\x 
         + g(\E+\v\times\B)\cdot\grad_\p\right] f
   = 0 ,
\label {eq:boltzmannA}
\end {equation}
where $\v$ is the velocity associated with $\p$, and $\E$ and $\B$ represent
the fields of the long-wavelength modes (those not described by $f$).
The corresponding Maxwell equations for the soft modes are
\begin {equation}
   \partial_\nu F^{\mu\nu} = j^\mu = g \int_\p v^\mu f ,
\end {equation}
where there is an implicit sum over particle species on the right-hand
side and $v^\mu \equiv (1,\v)$.
We use the $({-}{+}{+}{+})$ metric convention.
One can generalize these fully non-linear
Vlasov equations to non-Abelian plasmas.
However, for the discussion in this paper, it is simpler to
discuss the linearization of the equations in small fluctuations
$\delta f(\p,\x,t)$ of the distribution functions about
some initial homogeneous distribution $f_0(\p)$.
In this case,
\begin {subequations}
\label {eq:vlasovA}
\begin {equation}
   (\partial_t + \v\cdot\grad_\x) \, \delta f
          + g(\E+\v\times\B)\cdot\grad_\p f_0
   = 0 ,
\end {equation}
\begin {equation}
   \partial_\nu F^{\mu\nu} = j^\mu = g \int_\p v^\mu \, \delta f
\end {equation}
\end {subequations}
for the Abelian theory, assuming that $f_0$ carries no net charge or current.
The non-Abelian generalization which describes the dynamics of long
wave-length color electromagnetic fields is
\cite{mrowKinetic,HeinzKinetic,BIkinetic}
\footnote{
   Also see Ref.\ \cite{kelly} for a related formulation.
}
\begin {subequations}
\label {eq:vlasov}
\begin {equation}
   (D_t + \v\cdot \D_\x) \, \delta f
          + g(\E+\v\times\B)\cdot\grad_\p f_0
   = 0 ,
\label {eq:boltzmann}
\end {equation}
\begin {equation}
   D_\nu F^{\mu\nu} = j^\mu = c g \int_\p v^\mu \, \delta f ,
\label {eq:Maxwell}
\end {equation}
\end {subequations}
where $f_0$ is colorless and $\delta f$ takes values in the
adjoint color representation.  $D$ is the adjoint-representation
covariant derivative, and $c$ is a constant that depends on
the color representation of the hard particles.%
\footnote{
   Specifically, making the implicit sum over hard particle types $s$
   explicit, the right-hand side of (\ref{eq:Maxwell}) is
   $\sum_s c_s g \int_\p v^\mu \, \delta f_s$, where
   $c_s = \nu_s t_s = \nu_s C_s d_s/d_{\rm A}$.
   Here, $t_s$ is defined by $\tr(T^a_s T^b_s) = t_s \delta^{ab}$,
   where $T^a_s$
   are the color generators for the hard particle's color representation,
   $C_s$ is defined by $T^a_s T^a_s = C_s$, $d_s$ is the dimension of
   the color representation, and $d_{\rm A}$ is the dimension of the
   adjoint representation.  $\nu_s$ represents the number of non-color degrees
   of freedom of types $s$.
   For instance, for hard gluons in QCD, $d_s = 8$, $t_s=C_s=3$,
   and $\nu_s = 2$ for helicity provided $f$ represents the
   density of hard gluons {\it per}\/ spin state and color.
}
Note that we have linearized in $\delta f$ but {\it not}\/ in the
strength of the soft gauge field $A$.  This is the usual starting
point for kinetic theory discussions of hard thermal loops (HTLs)
\cite{braaten&pisarski}
in non-Abelian plasmas, whether for isotropic or anisotropic $f_0$.

There are a number of conditions that must be met for collisionless
Vlasov equations to give an accurate approximation to the underlying physical
situation.  The distance scales of interest must be (i) large compared to
the deBroglie wavelength of the hard particles, so that hard particle
propagation can be treated classically, and (ii) small compared to the
mean free path for individual hard particle collisions,
so that hard particle collisions may be ignored.
The soft fields must be large enough that they can be treated
as classical fields ({\it i.e.}\
the energy in each mode $k$ of interest should be large compared
to the frequency $\omega_k$ of that mode).  The last assumption will not
be a problem for a growing unstable mode, once it has grown to significant
size.

In the application to the thermalization of the quark-gluon plasma,
the ``hard particles'' are the gluons with momenta of order the
saturation scale $\Qs$.
The ``soft'' fields will refer to unstable
modes of the gauge fields with much smaller momentum.  (Note that hard
and soft are relative terms, and, in much of the literature on saturation,
the momentum scale $\Qs$ would normally be thought of as a ``soft'' scale.)
In order to keep the discussion of this paper somewhat general, we
will henceforth refer to the hard momentum scale as $p_\hard$ or simply
$p$ rather than $\Qs$, and we will refer to the soft momentum scale
as $k_\soft$ or simply $k$.
We will assume $k_\soft \ll p_\hard$.  This is the case for
the plasma instabilities of the original bottom-up scenario at times
$\tau \gg 1/\Qs$ \cite{alm}.

Though the precise value is not important to our discussion,
the constant $c$ in (\ref{eq:Maxwell}) is $c=2 C_{\rm A}=6$
for hard gluons if $f$
represents the density of hard gluons per helicity and color state.
$C_{\rm A}$ is the quadratic Casimir of the adjoint color
representation, with $C_{\rm A}=N$ for SU($N$).

We can now lay out more precisely
the basic question we will discuss in this paper.
There are two distinct natural scales associated with the question of
how big soft fields can grow before their effects can no longer be
treated perturbatively in Eqs.\ (\ref{eq:vlasov}).
The first has to do with the Boltzmann equation
(\ref{eq:boltzmann}) and with when
soft fields have a large effect on the trajectories of the
hard particles---that is, with when the assumption $\delta f \ll f_0$
breaks down.  Consider a soft magnetic field $B$.  The hard particles
will follow nearly straight-line ({\it i.e.}\ nearly free) trajectories if
the Larmor radius $R \sim p_\hard/gB$ is large compared to the
wavelength $\lambda \sim 1/k_\soft$ of the magnetic field.  That is,
the effects of the soft field on the hard particle trajectories
are perturbative if
\begin {equation}
   B \ll \frac{k_\soft \, p_\hard}{g} \,.
\label {eq:B}
\end {equation}
If one picks a ``reasonable'' gauge, where the gauge fields $A$ are
relatively smooth and do not have variation on scales different from
$k_\soft$, then $B \sim k A$, and this condition can be restated as%
\footnote{
  A simple mnemonic for this condition is to think momentarily about
  the more fundamental gauge theory that describes the hard modes,
  and ask when the soft contribution to the gauge field $A$ in the
  covariant derivative $D = \partial - i g A$ can be treated as
  a perturbation when applied to the hard modes.  When applied to hard
  modes, $D \sim p_\hard - g A$, and so the condition is
  $g A \ll p_\hard$.
}
\begin {equation}
  A \ll \frac{p_\hard}{g} \,.
\end {equation}
In non-Abelian plasmas, however, we also have to think about the
interaction of soft modes with each other.  A quick way to estimate
when these are important is to consider the covariant derivative
$D = \partial - igA$ in the non-Abelian Maxwell equation
(\ref{eq:Maxwell}).  When applied to the soft field strength $F$,
we have $D \sim k_\soft - g A$.  The $gA$ term can only be treated
as a perturbation when $gA \ll k_\soft$, which is
\begin {equation}
  A \ll \frac{k_\soft}{g} \,.
\end {equation}
Which of these parametrically different scales determines how large
the instabilities grow?  Is it
\begin {equation}
  A \sim \frac{p_\hard}{g} \,,
\label{eq:boundA}
\end {equation}
giving $B \sim kp/g$,
or
\begin {equation}
  A \sim \frac{k_\soft}{g} \ll \frac{p_\hard}{g} \,,
\label{eq:boundNA}
\end {equation}
giving $B \sim k^2/g \ll kp/g$\thinspace?
That is, is the growth stopped by large effects on hard particle
trajectories, or stopped
earlier by non-Abelian interactions between the growing soft modes?

In the next section, we will briefly review the linearized theory of
plasma instabilities.
In section \ref{sec:potential}, we then evaluate the
effective potential for a certain class of unstable configurations
and argue that, when non-linear interactions of the soft modes are
considered, the instabilities will grow beyond the non-Abelian scale
(\ref{eq:boundNA}) by evolving toward Abelian configurations.
In section \ref{sec:numerics}, we check this assertion in a
1+1 dimensional toy model, which we simulate numerically.
We find that the growing instabilities indeed evolve into
gauge configurations living in the maximal Abelian sub-algebra
of the gauge theory, which is U(1)$\times$U(1) for QCD.
In section \ref{sec:fate}, we briefly review what is known about
the fate of plasma instabilities from studies of traditional
plasmas, based on U(1) electromagnetism, once they grow large enough
to saturate the Abelian bound (\ref{eq:boundA}).
As an aside, in section \ref{sec:bound} we point out that there is
a simple lower bound on the exponent $n$ in the parametric relation
(\ref{eq:taueq}) for the equilibration time.
Finally, we offer our conclusions in section \ref{sec:conclusion}.
A number of topics briefly discussed in the main text are expanded
upon in various appendices.


\section {Review of linear instability}

One may formally solve the linearized Boltzmann equation (\ref{eq:boltzmann})
for $\delta f$ and then insert the solution into Maxwell's equation
(\ref{eq:Maxwell})
to obtain an effective equation of motion for the soft gauge fields:
\begin {equation}
   D_\nu F^{\mu\nu}
   = - c g^2 \int_\p v^\mu
     (D_t + \v\cdot\D_\x + \epsilon)^{-1} (\E + \v\times\B)\cdot\grad_\p f_0 ,
\label {eq:motion}
\end {equation}
where $\epsilon$ is a positive infinitesimal that selects the retarded
solution.
This equation simplifies enormously if one specializes to small gauge
fields by linearizing in $A$.  This limit has been analyzed by many
authors, and the result is similar to that for an Abelian plasma of
charged particles.  Henceforth dropping the subscript on $f_0$,
linearization yields
\begin {equation}
   \partial_\nu F^{\mu\nu}
   \simeq - c g^2 \int_\p v^\mu
     (\partial_t + \v\cdot\grad_\x + \epsilon)^{-1}
     (\E + \v\times\B)\cdot\grad_\p f .
\end {equation}
Fourier transforming from $(t,\x)$ to $K = (\omega,\k)$, this can be put
in the form
\begin {equation}
   i K_\nu F^{\mu\nu} \simeq - \Pi^{\mu\nu} A_\nu
\end {equation}
or equivalently
\begin {equation}
   S^{\mu\nu} A_\nu = 0 ,
\label {eq:A}
\end {equation}
where
\begin {equation}
   S^{\mu\nu}(\omega,\k) \equiv
   (-\omega^2 + k^2) g^{\mu\nu} - K^\mu K^\nu +
          \Pi^{\mu\nu}(\omega,\k)
\label {eq:S}
\end {equation}
and
\begin {equation}
   \Pi^{\mu\nu}(\omega,\k)
   = c g^2 \int_\p \frac{\partial f(\p)}{\partial p^k}
     \left[
       - v^\mu g^{k\nu} + \frac{v^\mu v^\nu k^k}{v\cdot K - i\epsilon}
     \right] .
\label {eq:Pi}
\end {equation}
We will assume that $f(\p)$ is parity-invariant in what follows.

One can now investigate whether there are unstable solutions to
(\ref{eq:A}), meaning solutions where $\omega$ has a positive imaginary
part.  Ref.\ \cite{alm} uses continuity arguments to show that a
sufficient condition for the existence of such an instability
with a given wavenumber $\k$ is that the spatial part
$S_{ij}$ of $S(\omega,\k)$ have a negative eigenvalue
at zero frequency.  That is,
\begin {equation}
   \varepsilon_i
   \bigl[ k^2 \delta_{ij} - k_i k_j + \Pi_{ij}(0,\k) \bigr] \varepsilon_j
   < 0
\label {eq:penrose}
\end {equation}
for some spatial polarization ${\bm \varepsilon}$ (which may be assumed
transverse to $\k$ without loss of generality).
This is a generalization of what's known in plasma physics as the
Penrose criterion.
(See also the earlier discussion in Ref.\ \cite{mrow2}.)
Roughly speaking,
Ref.\ \cite{alm} showed that this condition is satisfied for some $\k$
when $f(\p)$ is anisotropic.
Note that a negative eigenvalue for $S_{ij}(0,\k)$ requires a negative
eigenvalue for the spatial self-energy $\Pi_{ij}(0,\k)$.

\begin{figure}
\includegraphics[scale=0.50]{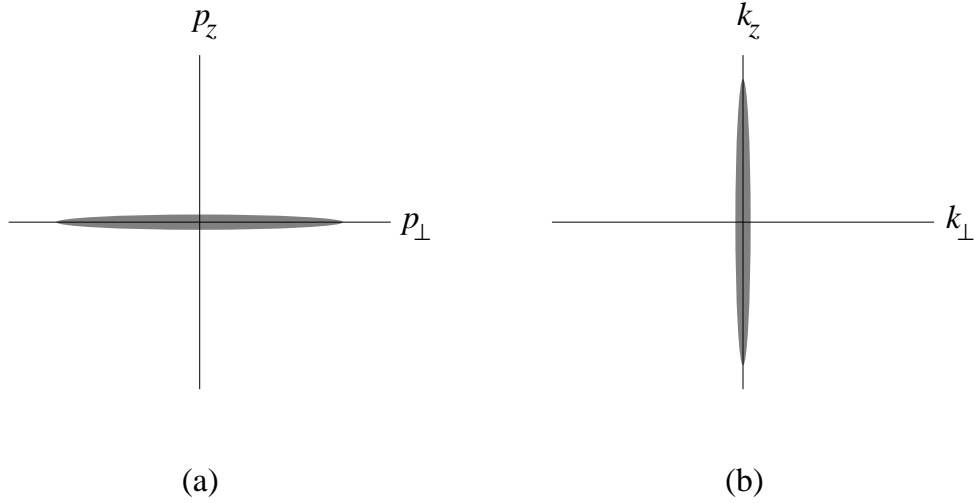}
\caption{%
    \label{fig:planar}
    (a) Flat pancake shape of the distribution of hard particle momenta
    $\p$ in the early stages of the original bottom-up scenario.
    (b) Associated narrow cigar shape of the set of soft momenta $\k$ of
    unstable modes of the soft gauge fields.  The scales in the two
    figures are not the same: $k_\soft \ll p_\hard$.
    }
\end{figure}

Ref.\ \cite{alm} also studied in detail a specific case relevant to
the early stages of the original bottom-up scenario, where
$f(\p)$ is axially symmetric and extremely anisotropic, with
typical $p_z \ll p_x, p_y$, as depicted schematically in
Fig.\ \ref{fig:planar}a.  Such a distribution has a flat pancake shape
in $\p$ space.  Qualitatively, the resulting set of soft, unstable modes
arising from (\ref{eq:A}) was found to look like the region of $\k$
space depicted in Fig.\ \ref{fig:planar}b, which has a narrow cigar shape
aligned along the $z$ axis.  In this particular situation, typical
unstable modes have $\k$'s which are almost parallel to the $z$ axis.
Whether such extreme anisotropy will persist in the theory
of quark-gluon plasma equilibration, once the effect of plasma instabilities
are fully incorporated into the bottom-up scenario, is not yet known.

Before moving beyond the linearized effective theory of the soft
gauge fields, it will be useful to note that the linearized equation
of motion (\ref{eq:A}) can be associated with a corresponding
effective action for the linearized theory:
\begin {equation}
   S_{\rm eff}^{\rm(linearized)}
   = \int_x
     \left[ - \fourth \, {\cal F}^a_{\mu\nu} {\cal F}^{a\mu\nu}
        - \half \, A^a_\mu \Pi^{\mu\nu} A^a_\nu \right] ,
\end {equation}
where $x$ stands for $(t,\x)$, and
${\cal F}_{\mu\nu} = \partial_\mu A_\nu - \partial_\nu A_\mu$
is the linear piece of the non-Abelian field strength $F_{\mu\nu}$.
Note that the action is non-local because $\Pi^{\mu\nu}$, given
by (\ref{eq:Pi}), is a non-local
operator in $x$.
In $A_0 = 0$ gauge, the action takes the form
\begin {equation}
   S_{\rm eff}^{\rm(linearized)}
   = \int_x
     \left[ \half |\partial_t \A|^2 - \half |\grad\times\A|^2
        - \half \, A_\mu \Pi^{\mu\nu} A_\nu \right] .
\label {eq:A0}
\end {equation}
Studying the condition (\ref{eq:penrose}) for instability is equivalent
to
looking for unstable directions of the ``potential energy''
defined by considering the effective action
(\ref{eq:A0}) for static (that is, $\omega{=}0$) gauge fields:
\begin {eqnarray}
   V[\A] &=& \int_\x \left[ \half |\grad\times\A|^2
            + \half A_i \, \Pi_{ij}(\omega{=}0) \, A_j \right]
\nonumber\\
         &=& \int_\k \half\, A_i(\k)^*
           \bigl[ k^2 \delta_{ij} - k_i k_j + \Pi_{ij}(0,\hat\k) \bigr]
           A_j(\k) .
\label {eq:Vlin}
\end {eqnarray}

In (\ref{eq:Vlin}), we have written $\Pi_{ij}(0,\hat\k)$ instead of
$\Pi_{ij}(0,\k)$.  This is because the formula (\ref{eq:Pi}) for
$\Pi(\omega,\k)$ depends only on the direction $\hat\k$ (and
not the magnitude) of $\k$ when $\omega = 0$.  We write
$\Pi(0,\hat\k)$ to emphasize this.
In the limit $k\to 0$, the instability condition that $S_{ij}(0,\k)$ have
a negative eigenvalue is equivalent to
$\Pi_{ij}(0,\hat\k)$ having a negative eigenvalue.%
\footnote{
  We should warn the reader that not all plasma instabilities of the
  system always correspond to negative eigenvalues of
  $\Pi_{ij}(0,\hat\k)$.  See the discussion of ``electric''
  instabilities in Ref.\ \cite{alm}.  However, for $\k$ pointing close
  enough to the $z$ axis for distributions
  like Fig.\ \ref{fig:planar}a,
  the negative eigenvalues of $\Pi_{ij}(0,\hat\k)$ do
  find all of the instabilities \cite{alm,strickland}.
} 

Before continuing, we should mention some limitations of the effective
potential.
Note that, in $A_0{=}0$ gauge, the action for a free electromagnetic field
$\int d^4x [\half |\partial_t \A|^2 + \half |\B|^2]$ looks like a kinetic
energy term $\half |\partial_t \A|^2$ which gives the electric energy,
plus a potential energy term $\half |\B|^2$ which gives the magnetic
energy.  Generally, the effective potential in $A_0=0$ gauge represents
magnetic potential energy, and electric effects are dynamical effects.
As a result, one does not see electric effects such as Debye screening
looking only at the effective potential (\ref{eq:Vlin}).  But one
can use the effective potential as a way of understanding instabilities
that manifest as negative eigenvalues of $\Pi_{ij}(0,\hat\k)$.


\section {The effective potential}
\label {sec:potential}

\subsection {Hard particle effects}

In order to investigate whether self-interactions of the soft gauge
fields can stop the growth of the instabilities, we wish to analyze
the effective potential energy {\it without}\/ linearizing the effective
theory in the gauge field $A^\mu$.
\Mrowczynski, Rebhan, and Strickland \cite{mrs} have derived the
effective action which gives the non-linear equation of motion
(\ref{eq:motion}).
Their action for anisotropic $f$ is a generalization of the
simple form of the HTL effective action
derived by Braaten and Pisarski
\cite{Sbp}%
\footnote{
   The HTL effective action for isotropic systems
   was originally derived, in a different form,
   by Taylor and Wong \cite{taylor&wong}.
   Ref.\ \cite{mrs} also discusses the generalization of the
   Taylor-Wong form to anisotropic systems.
}
for isotropic $f$.  It is given by%
\footnote{
   The difference in overall normalization of (\ref{eq:Seff1})
   with Ref.\ \cite{mrs},
   taking $c = 2 C_{\rm A}$ for hard gluons,
   is due to different choices of normalization for $f$.
}
\begin {equation}
   S_{\rm eff} =
   - \int_x \fourth {F^a}_{\mu\nu} F^{a\mu\nu}
   - c g^2 \int_x \int_\p
     \frac{f(\p)}{p} \, {F^a}_{\alpha\mu}(x)
     \left( \frac{v^\mu v^\nu}{(v\cdot D)^2} \right)_{ab}
     {{F^b}_\nu}^\alpha(x) .
\label {eq:Seff1}
\end {equation}
To get the potential energy, we evaluate this action for static
configurations $\A(\x)$ in $A_0=0$ gauge, as in the previous section.
Even with these restrictions, the formula (\ref{eq:Seff1})
remains a rather formal and complicated expression.
However, it simplifies tremendously if we restrict attention to
a certain sub-class of gauge field configurations
inspired by Fig.\ \ref{fig:planar}.
Because typical unstable $\k$'s in Fig.\ \ref{fig:planar}b have
$k_z \gg k_\perp$, generic combinations of these modes will vary
much more rapidly with $z$ than with $x$ or $y$.  Let us therefore
consider the extreme case of gauge field configurations that
depend only on $z$:
\begin {equation}
   \A = \A(z) .
\label {eq:Aformz}
\end {equation}
Making use of some observations by Blaizot and Iancu \cite{BIwaves},
we show in Appendix \ref{app:Veff} that, for $\A = \A(z)$,
the effective action (\ref{eq:Seff1}) reduces to the simple, local
form
\begin {eqnarray}
   V[\A(z)] &=& \int_\x \left[
      \fourth F^a_{ij} F^a_{ij}
       + \half A^a_i \, \Pi_{ij}(0,\hat\e_z) \, A^a_j \right]
\nonumber\\
   &=& \int_\x \left[
       \half \, \B^a\cdot\B^a
       + \half A^a_i \, \Pi_{ij}(0,\hat\e_z) \, A^a_j \right] ,
\label {eq:Vz}
\end {eqnarray}
where
$\Pi$ is the self-energy (\ref{eq:Pi}) of the linearized theory,
and $\hat\e_z$ is the unit vector in the $z$ direction.
$\B$ represents the full non-Abelian magnetic field.
This is exactly the same as the result (\ref{eq:Vlin}) of the linearized
theory applied to configurations $\A=\A(z)$
{\em except}\/ that the quadratic $|\grad\times\A|^2$ term
has been replaced by the full, non-Abelian $B^2$, which contains
cubic and quartic interactions.
The term $\A \Pi \A$ representing the effects of hard particles remains
quadratic, however, even though we are no longer linearizing the theory
in $\A$.

In field theory, one generally investigates questions of stability
by studying the effective potential for low-momentum modes, taking
$k \to 0$.  If we take this limit in (\ref{eq:Vz}), we obtain
the potential energy density
\begin {eqnarray}
   {\cal V} &=&
       - \fourth \, g^2 [A_i,A_j]^a [A_i,A_j]^a
       + \half A^a_i \, \Pi_{ij}(0,\hat\e_z) \, A^a_j ,
\nonumber\\
   &=&
       \fourth \, g^2 f^{abc} f^{ade} A^b_i A^c_j A^d_i A^e_j
       + \half A^a_i \, \Pi_{ij}(0,\hat\e_z) \, A^a_j ,
\end {eqnarray}
where $f^{abc}$ are the usual gauge-group structure constants.
We can now investigate the topography of this potential in the space
of $A^a_i$'s.
We will assume that the hard particle distribution functions $f(\p)$ are
axially symmetric about the $z$ axis and that $\Pi_{ij}(0,\hat\e_z)$
has a negative eigenvalue, which is the case for oblate
distributions like Fig.\ \ref{fig:planar}a.%
\footnote{
   For a qualitative discussion, see Ref.\ \cite{alm}.
   For calculations in various cases, see Refs.\
   \cite{randrup&mrow,strickland}.
}
Using the transversality $K_\mu \, \Pi^{\mu\nu}(K) = 0$ of the
HTL self-energy (\ref{eq:Pi}), we can then write
\begin {equation}
   {\cal V} =
       \fourth \, g^2 f^{abc} f^{ade} A^b_i A^c_j A^d_i A^e_j
       - \half \, \mu^2 (A^a_x A^a_x + A^a_y A^a_y) ,
\label {eq:pot}
\end {equation}
where
\begin {equation}
   \mu^2 \equiv -\Pi_{xx}(0,\hat\e_z) = -\Pi_{yy}(0,\hat\e_z) > 0.
\end {equation}
In the linearized analysis,
the unstable modes are those with $k < \mu$, and the typical
unstable modes have $k_\soft \sim \mu$ (that is, $k < \mu$ but typically
not $k \ll \mu$).

The potential (\ref{eq:pot}) is unbounded below.
Specifically,
consider any ``Abelian'' configuration, by which we mean a
gauge field where all the components $A_i$ of $\A$ commute.
(For instance, this happens if the gauge field points in a single
direction in adjoint color space, such as $A^a_i = A_i \delta^{a1}$.)
Then the quartic term in the potential vanishes, leaving
${\cal V} = - \half \, \mu^2 (A^a_x A^a_x + A^a_y A^a_y)$, which
clearly runs away to $-\infty$
as the magnitude of $A$ increases.
Note that this picture can no longer be trusted when $A$ gets
so large that the assumption $\delta f \ll f$ used to derive the
effective action breaks down.  That is, the runaway growth of Abelian
configurations should stop when the size of $A$ is given by the
scale $p_{\rm hard}/g$ of (\ref{eq:boundA}), as discussed in the introduction.

\begin{figure}
\includegraphics[scale=0.50]{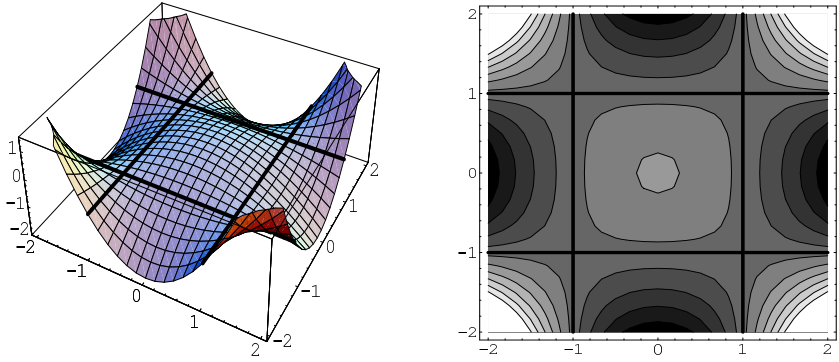}
\caption{%
    \label{fig:V}
    Two equivalent depictions of the potential $V(\phi_1,\phi_2)$
    of (\ref{eq:Vphi}).  The $\phi_1$ and $\phi_2$ axis are in units
    of $\mu/g$, and the values of $V$ are in units of $\mu^4/g^2$.
    There is a local maximum with ${\cal V}=0$ at the origin.
    The four straight lines in the plots correspond to
    the equipotential $V = -\mu^4/2g^2$, and
    the four intersection points of those lines
    are saddle points corresponding
    to static, unstable, non-Abelian configurations.
}
\end{figure}

We can get a simple picture of the topography of ${\cal V}$ for
non-Abelian configurations if we make some simplifying
restrictions on $\A$.  We will consider the case where both
(i) $\A$ lies in an SU(2) subgroup of color SU(3)
and (ii) $A_z = 0$.  With these assumptions, we show in Appendix
\ref{app:topo} that we can always make a combination of spatial and
color rotations to put $\A(k\to 0)$ in the following form:
\begin {equation}
   A_i^a = \phi_1 \delta_{ix} \delta_{a1} + \phi_2 \delta_{iy} \delta_{a2}
   .
\label {eq:Aform}
\end {equation}
The potential (\ref{eq:pot}) is then
\begin {equation}
   {\cal V}(\phi_1,\phi_2)  =
       \half \, g^2 \phi_1^2 \phi_2^2 - \half \, \mu^2 (\phi_1^2+\phi_2^2) .
\label {eq:Vphi}
\end {equation}
This potential is depicted in Fig.\ \ref{fig:V}.  The Abelian
configurations correspond to the $\phi_1$ axis ($\phi_2=0$)
and the $\phi_2$ axis ($\phi_1 = 0$).
One can see that there exist static non-Abelian solutions, indicated by
the intersection points of the four straight lines in the figure.
At these points, the amplitude of the gauge field is $A \sim \mu/g$.
Recalling that unstable modes typically have $k_\soft \sim \mu$,
this corresponds to the non-Abelian scale $A \sim k_\soft/g$ discussed in the
introduction.
However, these solutions are unstable to rolling down and
subsequently growing in amplitude along one of the axis.
The picture suggests that, if we start from $A$ near zero, the system
might possibly at first roll toward one of these configurations with
$A \sim k_\soft/g$, but its trajectory would eventually roll away and
approximately ``Abelianize,'' growing along either the $\pm \phi_1$ or
$\pm \phi_2$ axis until the effective action breaks down at
$A \sim p_\hard/g$.

We have been focusing on soft configurations $\A(z)$ in the arbitrarily long
wavelength limit $\k \to 0$.  In Appendix B, we show that
the static, unstable, non-Abelian solutions
discussed above have analogs with finite wavelength as well.
These new solutions are also unstable and are merely a curiosity: they
do not affect the current discussion.


\subsection {Soft mode effects}
\label{sec:hangup}

\begin{figure}
\includegraphics[scale=0.50]{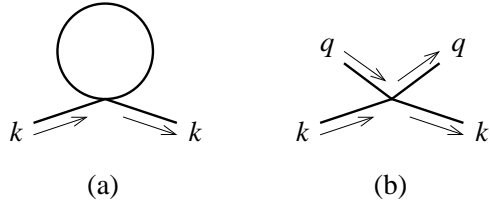}
\caption{%
    \label{fig:loop}
    (a) 1-loop mass correction in finite-temperature
    diagrammatic perturbation theory.
    (b) Forward scattering off of a particle in the plasma.
}
\end{figure}

So far, we have constructed the effective action and the effective potential
by integrating out the effects of the hard particles.
We have not yet, however, completely considered the effect that various
soft field excitations with different $\k$'s can have on the
instability.  In order to illuminate the remaining issues, let us consider
a very simple, warm-up toy model for the effective Lagrangian
for the soft fields,
which is a scalar field theory with the potential
${\cal V}(\phi_1,\phi_2)$ of (\ref{eq:Vphi}):
\begin {equation}
   {\cal L} = - \half (\partial\phi_1)^2 - \half (\partial\phi_2)^2
            - {\cal V}(\phi_1,\phi_2) .
\label {eq:ScalarToy}
\end {equation}
As an example of a concern one might have about our previous discussion
of instability, imagine that the soft sector were at some finite
temperature $T \gg \mu$.  What happens to our discussion when we account
for the effects of interactions with modes with momenta of order $T$?
It is well known that such interactions give a contribution
of order $g^2 T^2$
to the effective scalar mass-squared $m^2$.
This effective $\Delta m^2$ is generated
though diagrams such as Fig.\ \ref{fig:loop}a, which physically
represent the
forward scattering of particles off of the thermal bath, as in
Fig.\ \ref{fig:loop}b.
As a result,
\begin {equation}
   -\mu^2 \to -\mu^2 + O(g^2 T^2)
\end {equation}
in the effective potential (\ref{eq:Vphi}).  If $T$ were large enough,
this would stabilize the potential near the origin
and prevent runaway growth of the
fields.%
\footnote{
   Such a thermal mass effect would in fact only render the origin
   {\it meta}-stable in this scenario.
   The effective mass of small fluctuations behaves
   as $g\phi$ for large $\phi$ and exceeds $T$ for $\phi \gg T/g$.
   Thermal effects are then suppressed by $\exp(-g\phi/T)$,
   and so the stabilizing
   thermal contribution
   to the mass would disappear at large enough $\phi$.
}
This is the standard picture of high-temperature symmetry
restoration in scalar field theories \cite{linde,weinberg,dolan&jackiw}.

Can anything similar happen in the problem of interest, where we
consider an initial state that has very little energy in soft modes,
but these modes subsequently grow because of the instability?
In our toy scalar model, the answer is no.
For the scalar theory, we give simple qualitative arguments
(in any number of dimensions) in
Appendix \ref{app:qualitative}.
But a more straightforward way to verify our assertion is
by explicit numerical simulation, which we discuss in the next
section for 1+1 dimensions with a more sophisticated toy model
than the scalar theory (\ref{eq:ScalarToy}).


\section{Numerical analysis of Abelianization in a toy model}
\label{sec:numerics}

There are several uses for performing a numerical simulation
of a representative toy model of the physics we have been discussing.
One is simply to verify our claim in Sec.\ \ref{sec:hangup} that
soft excitations produced during instability growth cannot stabilize
the system from growth beyond the non-Abelian scale (\ref{eq:boundNA}).
Another is to better understand whether Abelianization is a global
phenomena.  We have suggested that the system, in searching to
minimize its potential energy as the fields grow large, seeks
configurations where commutators $[A_i,A_j]$ are small, since
these would otherwise contribute substantially to the non-Abelian
magnetic field energy $B^2$.  But is this a local statement or
a global statement?  Can one necessarily find a gauge where all
$A_i(\x)$ commute with all $A_j(\y)$ for well-separated
$\x$ and $\y$?  That is, over what distance scale can one say
that the non-Abelian gauge configurations are well approximated
by purely Abelian gauge configurations, and does this distance scale
grow with time?
In this section, we will address these questions by numerical simulations
of the real-time evolution of a 1+1 dimensional gauge theory.


\subsection {1+1 dimensional gauge theory toy model}

Consider the full effective theory (\ref{eq:Seff1}) of soft
modes but restrict attention to gauge fields of the form
\begin {equation}
   A_\mu = A_\mu(z,t) .
\end {equation}
This is similar to the restriction $A=A(z)$ of
the last section, except that we now allow for time dependence.
In {\it addition}, we will ignore time dependence in the
second term of (\ref{eq:Seff1}) so that it becomes
the simple local $\half A_i^a \Pi_{ij}(0,\hat\e_z) A^a_j$ of (\ref{eq:pot}).
This last approximation is not in general justified, which is why
this will only be a toy model calculation, and we will discuss some
of its deficiencies in Sec.\ \ref{sec:differences}.
We hope it will
qualitatively capture the physics of interest.
In 1+1 dimensional language, $A_0$ and $A_z$ are the gauge fields,
and $A_x$ and $A_y$ behave as adjoint-representation scalars.
To emphasize this distinction, we will refer to $A_x$ and $A_y$
as $\phi_x$ and $\phi_y$ in what follows.
The toy model Lagrangian is then
\begin {equation}
   {\cal L}
   =
   \half E_z^a E_z^a
   + \sum_{\alpha=x,y} \half [ (D_0 \phi_\alpha)^a (D_0 \phi_\alpha)^a
                  - (D_z \phi_\alpha)^a (D_z \phi_\alpha)^a ]
   - {\cal V}(\phi_x,\phi_y) ,
\label {eq:Ltoy}
\end {equation}
where ${\cal V}(\phi_x,\phi_y)$ is the potential of (\ref{eq:pot}).
By writing $\phi = T^a \phi^a$, where $T^a$ are fundamental representation
color generators with the usual normalization
$\tr(T^a T^b) = \half \delta^{ab}$,
we can also write this in the form
\begin {equation}
   {\cal L}
   =  \tr(E_z^2) 
      + \tr [ (D_0 \phi_x)^2 - (D_z \phi_x)^2 ]
      + \tr [ (D_0 \phi_y)^2 - (D_z \phi_y)^2 ]
      - {\cal V}(\phi_x,\phi_y) ,
\end {equation}
\begin {equation}
   {\cal  V}(\phi_x,\phi_y)
   = - \mu^2 \tr(\phi_x^2+\phi_y^2)
     + g^2 \tr \left(\bigl(i [\phi_x,\phi_y] \bigr)^2 \right)
   .
\end {equation}
In $A_0=0$ gauge, the equations of motion are
\begin {subequations}
\label {eq:eom}
\begin {eqnarray}
   \ddot \phi_1 &=& (D_z^2 + \mu^2)\phi_1 - g^2 [\phi_2,[\phi_2,\phi_1]] , 
\\
   \ddot \phi_2 &=& (D_z^2 + \mu^2)\phi_2 - g^2 [\phi_1,[\phi_1,\phi_2]] , 
\\
   \ddot A_z &=& i g [\phi_x, D_z \phi_x] + i g [\phi_y, D_z \phi_y] ,
\end {eqnarray}
\end {subequations}
with $D_z\phi = \partial_z\phi - i g [A_z,\phi]$.
Gauss's Law, which is a constraint equation preserved by the equations
of motion, is
\begin {equation}
   D_z \dot A_z = i g [\phi_x, \dot\phi_x] + i g [\phi_y, \dot\phi_y] .
\label {eq:gauss}
\end {equation}

Classical evolution does not depend in any essential way on the
value of $g$.
One can remove $g$ by a simple rescaling
of fields, $A_\mu \to A_\mu/g$ (including $A_x=\phi_x$ and $A_y=\phi_y$).
This is equivalent to setting $g=1$ in the equations (\ref{eq:eom}).
Recall that the typical unstable modes have momenta of order $\mu$.
It is therefore also natural to work in units where $\mu=1$, so that the
condition (\ref{eq:boundNA}) for when non-Abelian interactions between
growing unstable modes first becomes important is simply $A_\mu \sim 1$.
However, we will retain
factors of $g$ and $\mu$ in what follows simply so that our notation
and discussion is consistent with the conventions used earlier in this
paper.
Readers are encouraged to ignore these factors if so inclined,
setting $g=1$ and $\mu=1$ in what follows.

Our goal will be to evolve the system (\ref{eq:eom})
classically, starting from
small, random initial conditions for the fields.
Note that $A_0{=}0$ gauge is preserved by time-independent
gauge transformations.  In particular, in infinite volume,%
\footnote{
  We will be simulating finite volumes, where there can be
  non-trivial spatial Polyakov loops, but we use the infinite-volume
  case to inspire our choice of initial conditions.
}
one can always find a time-independent
gauge transformation to put the fields in axial gauge $A_z=0$ at
a particular time $t_1$.
We will use this freedom to choose our initial condition at $t=0$
to have $A_z=0$.  We want tiny initial fluctuations in our soft fields
to provide seeds for the soft-mode instabilities we wish to simulate.
For $A_x=\phi_x$ and $A_y=\phi_y$,
we choose their initial values to be zero and their
initial time derivatives to be Gaussian random white noise with
a very small amplitude $\Delta$.%
\footnote{
   Some readers may worry that using a white noise distribution may
   populate UV lattice modes with enough energy to seriously distort,
   through interactions,
   the physics of the soft sector in the continuum limit.
   This is not a problem and is discussed in Appendix \ref{app:lattice}.
}
Gauss's Law (\ref{eq:gauss})
then determines that $E_z = - \dot A_z$ must be $z$ independent at
$t=0$.  Since we are not interested in background electric fields,
we take $E_z$ to be initially zero.
In summary, our initial conditions (written in continuum notation) are
\begin {subequations}
\label {eq:initial}
\begin {equation}
   A_z(z,0) = \phi_x(z,0) = \phi_y(z,0) = 0 ,
  \qquad
   E_z(z,0) = 0 ,
\end {equation}
\begin {equation}
   \bigl\langle \dot\phi^a_\alpha(z,0) \, \dot\phi^b_\beta(z',0) \bigr\rangle
         = \frac{\Delta^2}{\mu} \,
           \delta^{ab} \delta_{\alpha\beta} \, \delta(z-z')
   .
\label {eq:white}
\end {equation}
\end {subequations}

In our simulations, we work on a 1-dimensional
periodic spatial lattice with lattice spacing $a = 0.05 \, \mu^{-1}$ and length
$L = 250 \, \mu^{-1} = 5000\,a$.
To evolve the system in time, we use
a leapfrog time algorithm with time step $\epsilon = 0.001\,\mu^{-1}$.
Our algorithm very closely follows that of
Refs.\ \cite{krasnitz,ambjorn&krasnitz}.%
\footnote{
  See also the overview in Ref.\ \cite{moore2} and the closely related
  algorithms discussed in Refs.\ \cite{grs,aaps}.
}
Our discrete-time lattice evolution equations, along with the
discretized version of our initial conditions, are given explicitly
in Appendix \ref{app:lattice}.
We choose $\Delta$, which parametrizes the size of the initial
fluctuations of $\dot\phi$, to be
$\Delta = 0.01 \, (a\mu)^{1/2} \mu^2/g \simeq 0.002236 \, \mu^2/g$.
We evolve a single, representative, random initial field from the
ensemble defined by (\ref{eq:white}).

In addition to simulating the model for SU(3) gauge theory, we have
also simulated it for SU(2) gauge theory.  The color structure of
SU(2) is slightly easier to discuss pedagogically because its
maximal Abelian subgroup is U(1).  If two adjoint fields commute in
SU(2), they must be pointing in the same color direction.
SU(3) is more complicated.
For this reason, we start by discussing SU(2).


\subsection {Simulation Results: SU(2) gauge theory}

\begin{figure}
\includegraphics[scale=0.40]{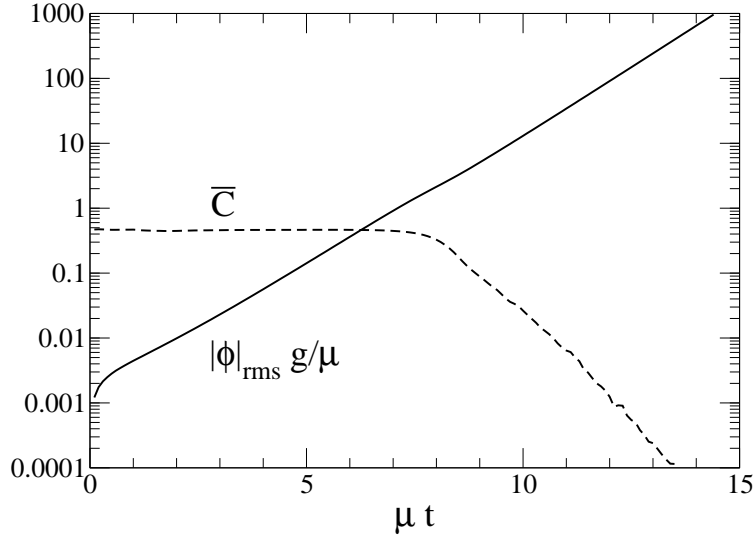}
\caption{%
    \label{fig:SU2phi}
    The average amplitude $|\phi|_{\rm rms}$ (solid) and the relative
    size $\overline{C}$ of commutators (dashed) as a function of time for
    gauge group SU(2).
    }
\end{figure}

Fig.\ \ref{fig:SU2phi} shows how the amplitude of the fields grows with
time.  Specifically, it shows the growth of the rms average
\begin {equation}
   |\phi|_{\rm rms} \equiv
   \left[
      \int_0^L \frac{dz}{L} \> (\phi_x^a \phi_x^a + \phi_y^a \phi_y^a)
   \right]^{1/2}
   = \left[
      \int_0^L \frac{dz}{L} \> 2 \tr(\phi_x^2 + \phi_y^2)
   \right]^{1/2} .
\end {equation}
The figure also shows the volume average of the
relative size of the local commutator $[\phi_x,\phi_y]$ of the
fields $A_x = \phi_x$ and $A_y = \phi_y$, as measured by
\begin {equation}
  \overline{C} \equiv
  \int_0^L \frac{dz}{L} \> C(z) ,
  \qquad
  C  \equiv
     \frac{ \left\{\tr \left(
                \bigl(i [\phi_x,\phi_y] \bigr)^2
            \right)\right\}^{1/2}
          }
          { \tr(\phi_x^2+\phi_y^2) }
     \, .
\label {eq:C}
\end {equation}

As one can see, there is a stage of the instability growth, around
$\phi \sim \mu/g$, where the original, random, non-Abelian fluctuations
suddenly change character and $\phi_x$ and $\phi_y$ become,
at least locally, aligned in the same color direction.  We will
analyze soon how global is this alignment.

Before continuing, we should note that quantities such as
$\tr(A_x^2 + A_y^2)$ and $\tr([A_x,A_y]^2)$ are gauge-invariant under
1+1 dimensional gauge transformations, but they would not be
be gauge-invariant in the original 3+1 dimensional theory under
general 3+1 dimensional gauge transformations.  This makes the
construction of gauge-invariant observables easier in the 1+1 dimensional
theory than in 3+1 dimensions.  In particular, we can directly probe
statements about the relative size of the commutator $[A_x,A_y]$
without having to either
(i) rely on the vague, qualitative statements like
``in a reasonable gauge''
which we used earlier in this paper, or (ii) construct more indirect
observables, like commutators of magnetic fields.
We shall take advantage of this feature of the 1+1 dimensional theory.

We have shown what happens to the relative size of $[\phi_x,\phi_y]$.
Readers may wonder about the other possible commutators of the
basic fields of our model, $[A_z, \phi_x]$ and $[A_z, \phi_y]$.
The value of $[A_z, \phi_\alpha]$ at a particular moment is not physical
because we can make it vanish everywhere by gauge transforming so that
$A_z=0$ at that moment.  We can think of fixing axial gauge
at a particular time $t$ as the 1+1 dimensional analogue of
a ``smooth'' gauge choice for that time slice.

\begin{figure}
\includegraphics[scale=0.40]{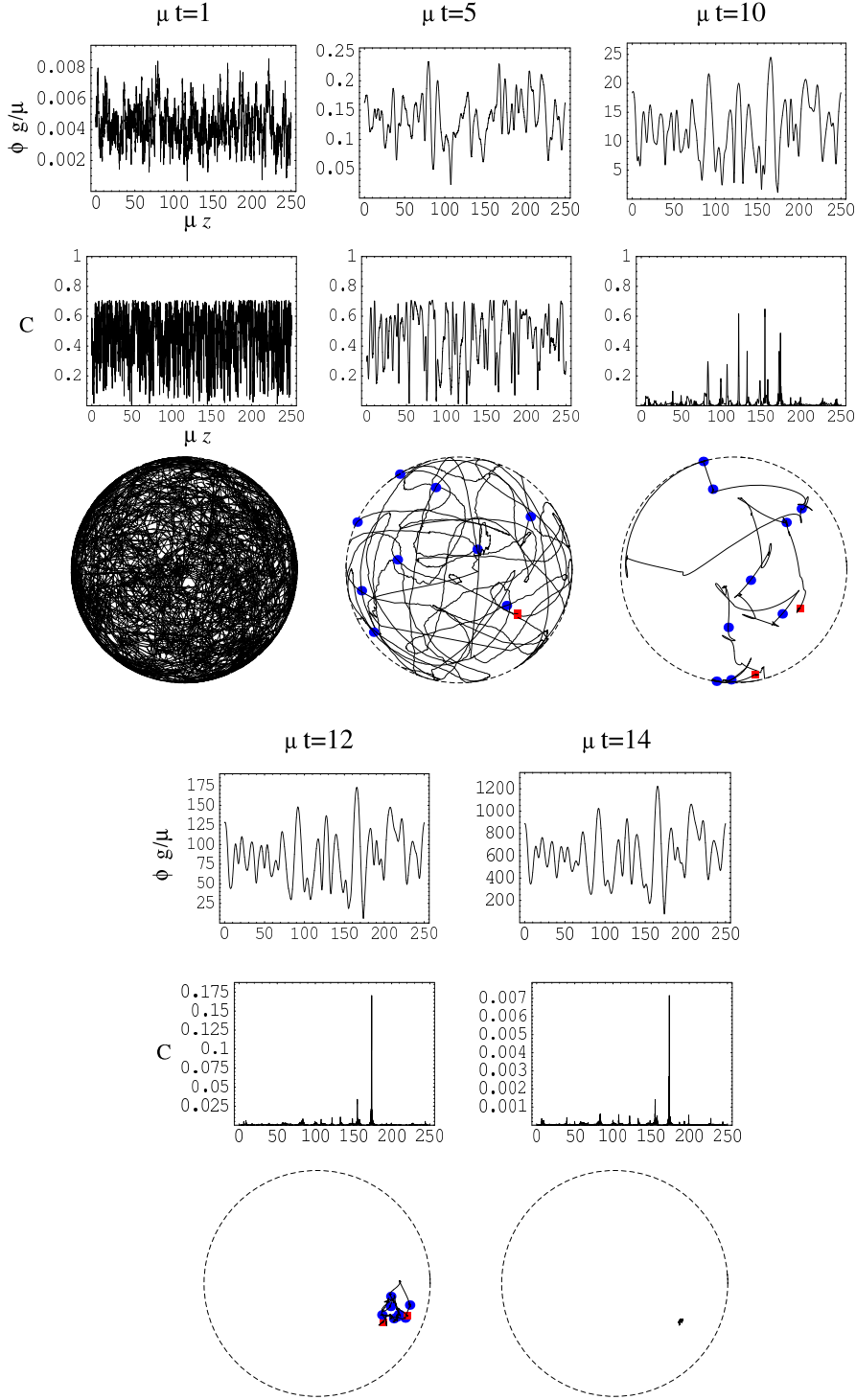}
\caption{%
    \label{fig:SU2array}
    Snapshots in time of the evolution of SU(2) field configurations.
    Each column corresponds to a different time, labeled at the top:
    $\mu t =$ 1, 5, 10, 12, 14.
    For each time, the first row shows the total field
    strength (\ref{eq:Phi}), in units of
    $\mu/g$, as a function of $\mu z$.
    The second row shows the relative size $C$ of commutators
    as a function of $\mu z$.
    The third row shows the the color directions
    swept out in axial gauge as one varies $z$ from $0$ to $L$, as
    described in the text.  The circles mark the points on these curves
    where $\mu z$ is a multiple of 50, and squares indicate the endpoints
    $z=0$ and $z=L$.
}
\end{figure}

Snapshots of the field configurations are shown at five different times
in Fig.\ \ref{fig:SU2array}.  The first row shows the field strength
\begin {equation}
   \Phi \equiv
    (\phi_x^a \phi_x^a + \phi_y^a \phi_y^a)^{1/2}
    = \left[ 2 \tr (\phi_x^2 + \phi_y^2) \right]^{1/2}
\label {eq:Phi}
\end {equation}
as a function of $z$.  The second row shows the relative size $C$ of
commutators, as defined in (\ref{eq:C}), again as a function of $z$.
Note that the scales used in the graphs change with time.

We have learned that $\phi_x$ and $\phi_y$ point in the same color
direction at late times.  We would now like to address whether
this direction ``changes'' with $z$.  That is, is it possible to find
a gauge where the fields all live in an Abelian sub-algebra of SU(2),
{\it e.g.} all pointing in the $T^3$ direction:
\begin {equation}
   A^a_z(z,t) = A_z(z,t) \, \delta^{a3} ,
   \qquad
   \phi^a_\alpha(z,t) = \phi_\alpha(z,t) \, \delta^{a3} .
\label {eq:Abelian}
\end {equation}
A gauge-invariant way to compare the color directions of $\phi_\alpha(z)$
for different $z$ is to parallel transport all of the $\phi_\alpha(z)$
to some reference point $z_0$.  We will pick $z_0=0$.  Equivalently, one
can simply gauge transform to $A_z{=}0$ gauge on the particular time slice
of interest, in which case parallel transport is trivial and the color
directions at different $z$ can be compared directly.

\begin{figure}
\includegraphics[scale=0.50]{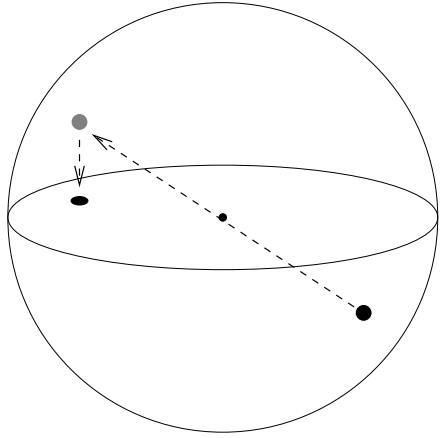}
\caption{%
    \label{fig:unitS2}
    Projection of the space $S^2$ of SU(2) adjoint
    color directions into the plane, with the identification of
    antipodal points.
}
\end{figure}

In SU(2), the color direction of an adjoint field $\phi^a$ can be
represented as a unit 3-vector
$(\phi^1,\phi^2,\phi^3)/(\phi^a\phi^a)^{1/2}$.
It therefore lives on a 2-sphere $S^2$.
However, the fields of an Abelian configuration (\ref{eq:Abelian})
can take both positive and negative values, corresponding, for example,
to color directions $(0,0,\pm 1)$.  So, to test how much a given
configuration deviates from being Abelian, we should not differentiate
between a given color direction and its negative.  The relevant color
space is therefore $S^2$ with antipodal points identified ($S^2/Z_2$).
The third row of Fig.\ \ref{fig:SU2array} shows the (parallel transported%
\footnote{
  In a periodic space, there are two different ways to parallel transport
  to $z=0$: to the left or to the right.  Our convention in these plots is
  to always parallel transport to the left.  Equivalently,
  there may be an obstruction to transforming to $A_z=0$ gauge
  everywhere in a periodic space.
  Our convention here is to transform to $A_z=0$ everywhere
  except on the link that connects across the periodic boundary that
  identifies $z=0$ and $z=L$.
})
color directions swept out as $z$ varies from $0$ to $L$.
The color direction%
\footnote{
   \label{foot:dir}
   Readers may wonder which color directions we have plotted, since there
   are two different fields, $\phi_x$ and $\phi_y$.  For an Abelian field,
   ${\bm\phi}\equiv(\phi_x,\phi_y)$ points in a single, well-defined direction
   ${\bm\varepsilon}(z,t)$
   in the $xy$ plane at each point.
   For general non-Abelian fields, we simply define
   ${\bm\varepsilon}$ as the unit vector which maximizes
   $({\bm\varepsilon\cdot\bm\phi})^a ({\bm\varepsilon\cdot\bm\phi})^a$.
   The plots in the third row of Fig.\ \ref{fig:SU2array} are of
   the color direction of ${\bm\varepsilon\cdot\bm\phi}$.  At late times,
   this aligns with the color directions of $\phi_x$ and $\phi_y$
   except at those isolated points where ${\bm\varepsilon}$ is perpendicular to
   $x$ or $y$ (where the very tiny non-Abelian components of
   $\phi_x$ or $\phi_y$ will dominate over the Abelian one, which is
   simply an accident of how one chooses the $x$ and $y$ axis).
}
is represented in the following way.
Consider a color direction represented as a point on the unit sphere, as shown
in Fig.\ \ref{fig:unitS2}.  If it is in the lower hemisphere, use the
identification of antipodal points to replace it by a point in the
upper hemisphere.  Then project the points in the upper hemisphere down
to the plane passing through the equator.
The plots in Fig.\ \ref{fig:SU2array}
show that plane.
The dashed circles represent the equator, for which
antipodal points are identified.

\begin{figure}
\includegraphics[scale=0.50]{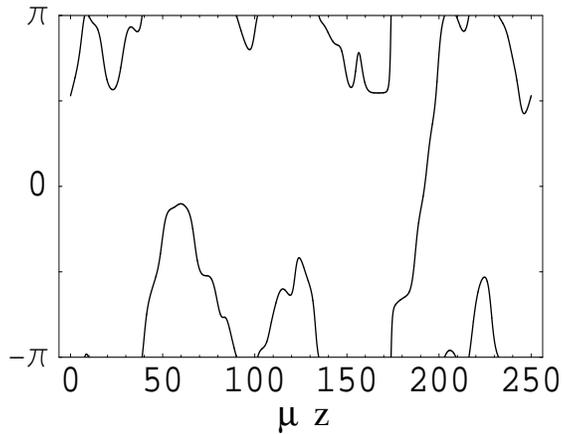}
\caption{%
    \label{fig:SU2angle14}
    The angle in the $xy$ plane of the vector
    ${\bm\phi} \equiv (\phi_x,\phi_y)$, as a function of $\mu z$
    at
    $\mu t = 14$, for SU(2) simulations
    (taking ${\bm\phi}$ projected
    onto the color direction indicated
    in the last plot of Fig.\ \ref{fig:SU2array}).
}
\end{figure}

One can see from the color direction plots in Fig.\ \ref{fig:SU2array} that
the color directions globally align as $t$ gets large.  At $\mu t = 14$,
the configuration is to very good approximation Abelian over
the entire simulation volume.  However, other aspects of the field remain
uncorrelated.  Fig.\ \ref{fig:SU2angle14} shows the angle corresponding
to the direction in the $xy$ plane of the essentially Abelian
${\bm\phi} \equiv (\phi_x,\phi_y)$ at $\mu t=14$, as a function of $z$.%
\footnote{
  More precisely, we plot the direction ${\bm\varepsilon}$ defined in
  footnote \ref{foot:dir}.
}
So, the SU(2) configurations at late times are homogeneous in color
and look like a random superposition of unstable Abelian modes.

\subsection {Abelianization Length: results for SU(2)}

The adjoint color space of SU(3) is more complicated than that of
SU(2).  When we study SU(3) in the next section,
we will not be able to make simple, visual plots of color
direction as in Fig.\ \ref{fig:SU2array}.
Also,
the maximal Abelian subgroup of SU(2) is U(1), but the maximal Abelian
subgroup of SU(3) is U(1)$\times$U(1).
In SU(3), colors
need not point in a single color direction in order to be Abelian.
It will therefore be useful
to replace our pictorial analysis by an appropriate correlation length
that measures over what distance scales fields commute with each other.
Consider the following correlation, which measures whether fields
commute over a given distance $\xi$:
\begin {multline}
   \chi_{\rm A}(\xi) \equiv
\\
   \frac{d_{\rm A}}{2C_{\rm A}}
   \int_0^L \frac{dz}{L}
   \frac{
     \tr \Bigl\{ i
     \bigl[\phi_\alpha(z+\xi), \, {\cal U}(z+\xi,z) \, \phi_\beta(z)\bigr]
     \, i
     \bigl[\phi_\alpha(z+\xi), \, {\cal U}(z+\xi,z) \, \phi_\beta(z)\bigr]
     \Bigr\}
   }{
     \tr\left\{\phi_\alpha(z+\xi) \, \phi_\alpha(z+\xi) \right\} \,
     \tr\left\{\phi_\beta(z) \, \phi_\beta(z) \right\}
   }
   \,,
\label {eq:chiA}
\end {multline}
where
${\cal U}(z',z)$ represents adjoint-representation parallel transport
from $z$ to $z'$.  Specifically,
\begin {equation}
   {\cal U}(z',z) \, \phi_\beta(z)
   = U(z',z) \, \phi_\beta(z) \, \bigl[ U(z',z) \bigr]^\dagger,
\end {equation}
where $U(z',z)$ is the fundamental-representation transporter
\begin {equation}
   U(z',z) = {\cal P} \exp\left[i g \int_z^{z'} dz'' \> A_z(z'') \right] ,
\end {equation}
and ${\cal P}$ indicates path ordering (with $z'$ on the left and
$z$ on the right).
Alternatively, one may transform to $A_z{=}0$ gauge and dispense with
the parallel transporters.%
\footnote{
   There is a subtlety to this on the lattice when $z < L < z+\xi$ because
   one cannot fix $A_z{=}0$ gauge on one link, which we choose as the
   link across the periodic boundary.  One can either incorporate this
   single link in the parallel transport, or else extend the axial
   gauge transformation to the periodic copies of the fields beyond $z=L$.
}
In (\ref{eq:chiA}), $d_{\rm A}$ and $C_{\rm A}$ are the dimension and
quadratic Casimir of the adjoint color representation.

\begin{figure}
\includegraphics[scale=0.40]{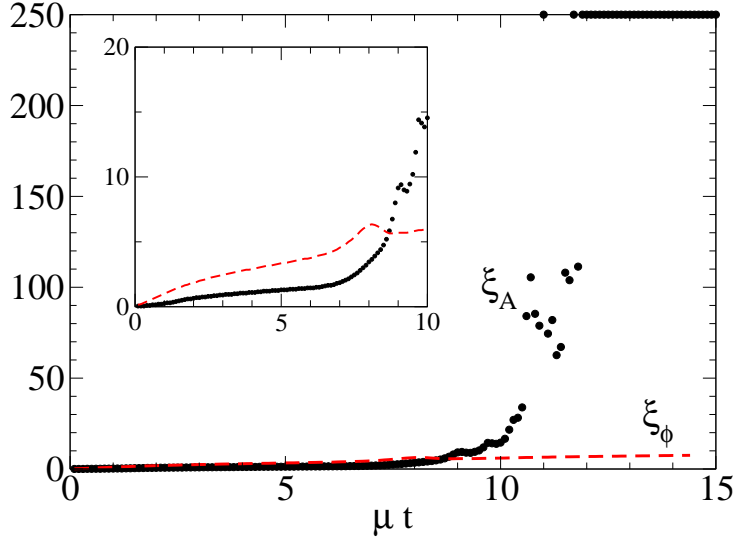}
\caption{%
    \label{fig:SU2correlation}
    The solid lines and circles show the Abelianization correlation length
    $\xi_{\rm A}$ as a function of time, measured every
    $\delta t = 0.1 \, \mu^{-1}$.  Individual points, rather than a continuous
    line, are shown simply because $\xi_{\rm A}$ jumps around quickly once
    it becomes a sizable fraction of the lattice.
    The dashed line shows the ordinary correlation length $\xi_\phi$.
    The topmost axis of the outer graph is $L/2$ and represents
    correlation lengths which exceed the size of our lattice.
    The inset provides an expanded view of the behavior for $\mu t < 10$.
}
\end{figure}

If commutators vanish locally, as we have seen they do for late times, then
$\chi_{\rm A}(0) = 0$.  The correlation
$\chi_{\rm A}$ has been normalized so that if the colors are
completely uncorrelated over a distance $\xi$, then $\chi(\xi)=1$.
We define the ``Abelianization correlation length'' $\xi_{\rm A}$
as the smallest
distance $\xi$ for which $\chi_{\rm A}(\xi) \ge 0.5$.
A plot of $\xi_{\rm A}$ vs.\ time for SU(2) is shown in
Fig.\ \ref{fig:SU2correlation}.  When $\xi_{\rm A}$ exceeds $L/2$
(indicated by the top of the plot), then the correlation does not
drop below 0.5 anywhere on the lattice.  One can see that this correlation
length begins to grow rapidly at roughly the same time the relative
size $\bar C$ of commutators begins to drop in Fig.\ \ref{fig:SU2phi}.

For comparison, we also plot in Fig.\ \ref{fig:SU2correlation}
a correlation length defined in terms of
the {\it full}\/ correlation (not just color) between the
parallel-transported
fields, defined by
\begin {equation}
   \chi_{\phi}(\xi) \equiv
   \frac{
      \int_0^L \frac{dz}{L} \tr\left\{
        \phi_\alpha(z+\xi) \, {\cal U}(z+\xi,z) \, \phi_\alpha(z) \right\}
   }{
      \int_0^L \frac{dz}{L} \tr\left\{
        \phi_\alpha(z) \, \phi_\alpha(z) \right\}
   } \,.
\end {equation}
This correlation is normalized so that $\chi(0)=1$, and $\chi(\xi)$
vanishes if the fields are uncorrelated over distance $\xi$.
We define a correlation length $\xi_\phi$ by when this correlation
first drops below 0.5.  Note that this correlation length does not grow
enormously like the Abelianization length $\xi_{\rm A}$ does.


\subsection {Simulation Results: SU(3) gauge theory}

\begin{figure}
\includegraphics[scale=0.40]{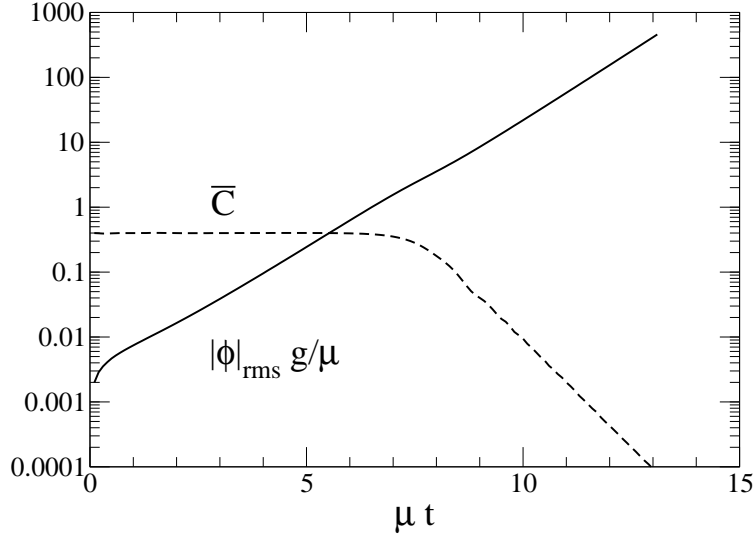}
\caption{%
    \label{fig:SU3phi}
    As Fig.\ \ref{fig:SU2phi} but for SU(3) gauge theory.
    }
\bigskip\medskip   
\end{figure}

\begin{figure}
\includegraphics[scale=0.40]{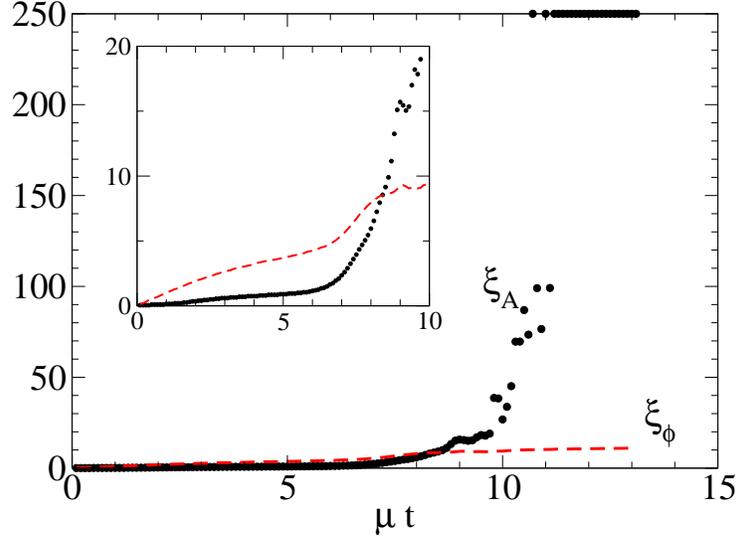}
\caption{%
    \label{fig:SU3correlation}
    As Fig.\ \ref{fig:SU2correlation} but for SU(3) gauge theory.
}
\end{figure}

We are now ready to discuss simulation results for the QCD version of
our 1+1 dimensional model.  Simulation results for $|\phi|_{\rm rms}$,
$\bar C$, $\xi_{\rm A}$ and $\xi_\phi$ for SU(3) are shown as a function
of time in Figs.\ \ref{fig:SU3phi} and \ref{fig:SU3correlation}.
The results are qualitatively similar to the SU(2) case, and we can
draw the same conclusion from the rapid growth of $\xi_{\rm A}$:
the configurations Abelianize as the field strength grows large compared
to the natural non-Abelian scale $A \sim \mu/g$.

Plots of $\Phi$ and $C$ vs.\ $z$ are qualitatively similar to those
of the SU(2) simulations shown in the first two rows of
Fig.\ \ref{fig:SU2array}, and we will refrain from displaying them.

\begin{figure}
\includegraphics[scale=0.40]{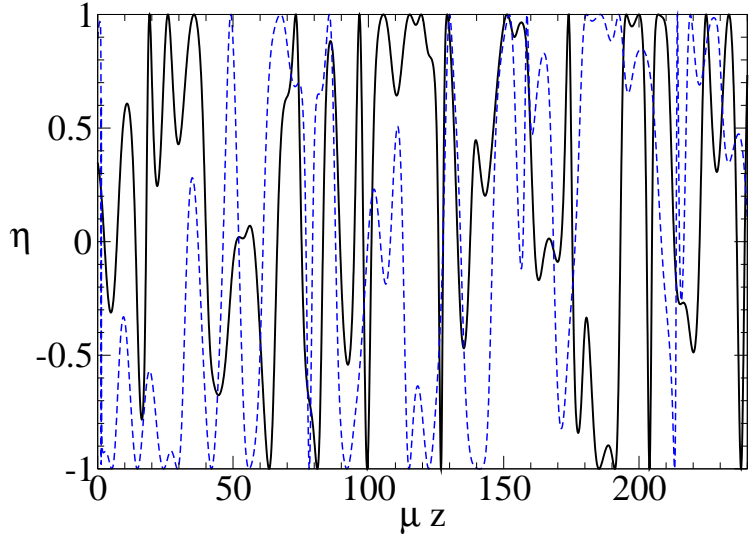}
\caption{%
    \label{fig:eight}
    The parameters $\eta_x$ (solid) and $\eta_y$ (dashed)
    of (\ref{eq:eta}) vs.\ $z$ for $\mu t = 13$,
    showing the spatial variation of color
    within the maximal Abelian subgroup of SU(3).
}
\end{figure}

The Abelianization in SU(3) is slightly different than in SU(2) because
the fields do not have to point in a single color direction.
The maximal Abelian subgroup of SU(2) is U(1)$\times$U(1) and,
after an appropriate gauge transformation, can be considered to be spanned
by the diagonal generators
\begin {equation}
   T^3 = \frac12 \begin {pmatrix}
            1 & & \\ & -1 & \\ & & 0 
         \end {pmatrix} ,
   \qquad
   T^8 = \frac1{\sqrt{12}} \begin {pmatrix}
            1 & & \\ & 1 & \\ & & -2 
         \end {pmatrix} ,
\end {equation}
in the standard Gell-Mann
representation.  The analog of (\ref{eq:Abelian}) for Abelian configurations
is that they can be put in the form
\begin {equation}
   A^a_z(z,t) = A^3_z(z,t) \, \delta^{a3} + A^8_z(z,t) \delta^{a8},
   \qquad
   \phi^a_\alpha(z,t) = \phi^3_\alpha(z,t) \, \delta^{a3}
                      + \phi^8_\alpha(z,t) \, \delta^{a8}.
\end {equation}
To investigate whether the late-time SU(3) configurations vary in color
with $z$ within the maximal Abelian subgroup, we would like to
plot the relative $T^3$ and $T^8$ components as a function of $z$.
$T^3$ and $T^8$ are somewhat artificial choices, however, because they break
the permutation symmetry of the U(1)$\times$U(1) subgroup, treating the
first two colors of the fundamental representation differently from the third.
In Fig.\ \ref{fig:eight}, we plot instead the gauge-invariant
measures
\begin {equation}
   \eta_\alpha(z) =
   \frac{3\sqrt{6} \, \det \phi_\alpha(z)}
        {\bigl[\tr(\phi_\alpha(z) \, \phi_\alpha(z))\bigr]^{3/2}}
   \qquad\qquad
   \mbox{(no sum on $\alpha$)}
\label {eq:eta}
\end {equation}
vs.\ $z$.  These parameters vary in magnitude $|\eta|$
between (i) zero, for color matrices with
eigenvalues proportional to $(1,-1,0)$, like $T^3$ and its permutations,
and (ii) 1,
for color matrices with eigenvalues proportional to $(1,1,-2)$, like $T^8$
and its permutations.  A plot of $\eta_\alpha(z)$ vs. $z$ is given for
time $\mu t = 13$.  As one can see, the color direction within the
maximal Abelian subgroup remains uncorrelated at this late time,
even though the Abelianization correlation length is larger than the
size of the lattice.

\subsection{Differences between the toy model and the full effective theory}
\label {sec:differences}

\begin{figure}
\includegraphics[scale=0.40]{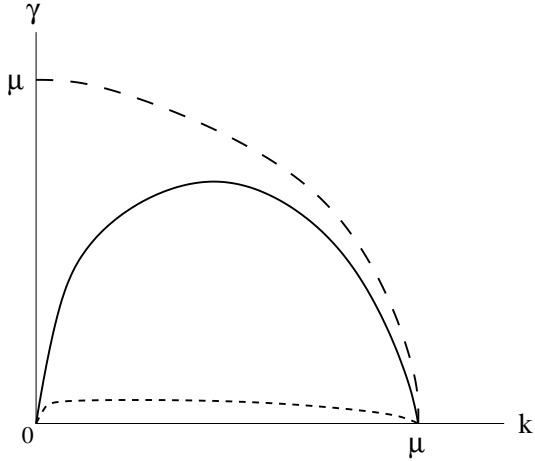}
\caption{%
    \label{fig:growth}
    Qualitative behavior of linear growth rates of instabilities vs.\
    momentum $k$ for (long-dashed) the toy model, (solid) actual QCD
    with a generic axi-symmetric hard particle distribution $f(\p)$,
    and (short-dashed) actual QCD with a pancake-shaped hard particle
    distribution.
}
\end{figure}

Before leaving our discussion of simulation results, we should mention
one of the qualitative differences between our toy model (\ref{eq:eom})
and the real theory of plasma instabilities.  In constructing the toy
model, we ignored time dependence in the second term of the
effective action (\ref{eq:Seff1}).  Consider a linear analysis of
instabilities around zero field.  In our toy model, the growth rate of
each mode with $k < \mu$ is given by
\begin {equation}
   \gamma_{\rm toy}(k) = \sqrt{\mu^2 - k^2} ,
\end {equation}
which is easily derived by linearizing the equations of motion
(\ref{eq:eom}).
Qualitatively, this growth rate has the form of the long-dashed line in
Fig.\ \ref{fig:growth} and is $\gamma_{\rm toy}=\mu$ at $k=0$.
In the actual theory, where time dependence is not ignored,
$\gamma(k)$ vanishes at $k=0$
\cite{mrow2,randrup&mrow,alm,strickland}.
For generic hard particle distributions, $\gamma(k)$ then has the
qualitative form of the solid line in Fig.\ \ref{fig:growth}.
The slow response of the system
for small $k$ is related to Lenz's Law:
changing magnetic fields create electric fields, which induce currents
to oppose the change.
For the extremely anisotropic case of pancake distributions,
such as shown in Fig.\ \ref{fig:planar},
instability growth is relatively slow compared to $\mu$
for all unstable $k$ \cite{alm}.%
\footnote{
  This follows from the results of Ref.\ \cite{alm} with the realization
  that $m_\infty \ll q_{\rm max}$ in that reference when the hard particle
  distribution is extremely pancake shaped.  The $q_{\rm max}$ of
  Ref.\ \cite{alm} is the $\mu$ of this paper.
}
This is depicted by the short-dashed line
in Fig.\ \ref{fig:growth}.
We find it plausible that slow down of the growth rate will not affect
the qualitative conclusions we have drawn: that energetics favors
the development of Abelian configurations.
However, this is something that should be checked in the future by
simulations of the full theory.
In particular, one might wonder whether the suppression of the growth
of small $k$ modes in the full theory might affect the growth of
the color correlation length to scales large compared to $\mu^{-1}$.

\section{Fate of Abelian instabilities}
\label {sec:fate}

If QCD plasma instabilities indeed Abelianize,
then we may be able to learn qualitative lessons about their behavior
from studies of traditional electromagnetic plasmas.  Indeed, for
SU(2) gauge theory, we have seen in our toy model that plasma instabilities
grow into configurations of U(1) gauge theory.
For QCD, the situation is a little more complicated: we get two copies
of ``electromagnetism'' as U(1)$\times$U(1).  Still, we may hope to
gain some qualitative insight from what's known about U(1) gauge theories.
In this section, we will review a few relevant results from the
traditional plasma literature.

\begin{figure}
\includegraphics[scale=0.40]{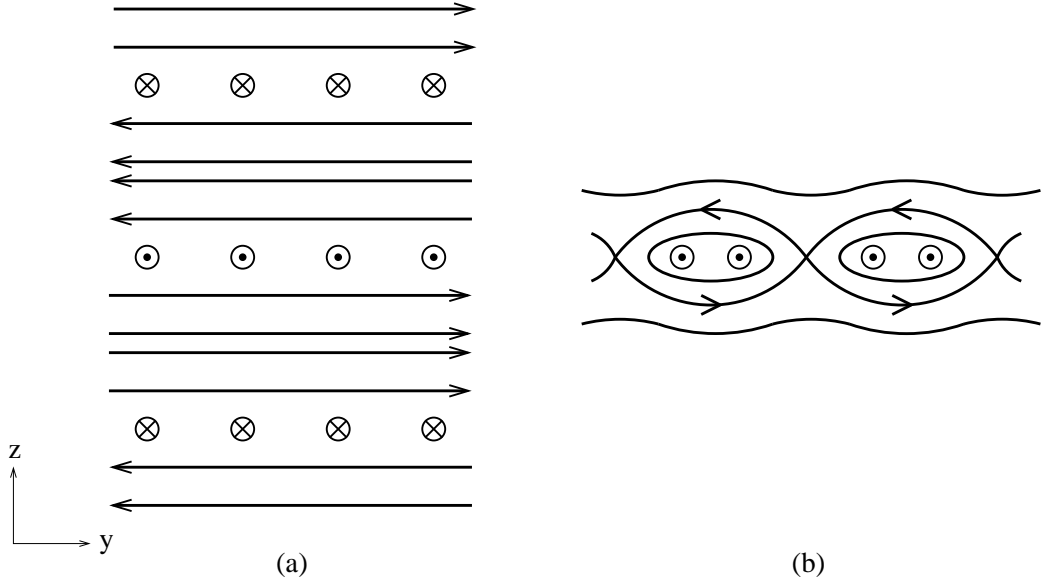}
\caption{%
    \label{fig:tear}
    (a) Qualitative depiction of a non-linear magnetic BGK wave.
        The lines represent magnetic field lines.  The points and
        crosses represent the net charged particle current, pointing
        either out of or into the page.
    (b) A magnetic tear or reconnection instability for the central
        region of figure (a).
}
\end{figure}

Throughout this paper, we have focused on configurations which are
approximately independent of $x$ and $y$, which arise naturally,
for example, when the distribution of hard particles is pancake-shaped.
One might wonder whether the dynamics continues to maintain the
approximate independence of the configurations on $x$ and $y$ once
the instabilities grow large enough that their effects on the hard
particles become non-perturbative.
Could the system settle down into oscillations about some
non-linear $z$-dependent magnetic configurations, such as shown
in Fig.\ \ref{fig:tear}a?
Researchers have found examples of time-independent, non-perturbative
wave solutions of the full (Abelian) non-linear collisionless Vlasov
equations, such as the magnetic Bernstein-Greene-Kruskal
(BGK) wave solutions
discovered by Davidson {\it et al.}\ \cite{davidson&etal,berger&davidson}.
However, one might guess that such solutions are inflicted with
yet another type of plasma instability, known in different contexts
as magnetic tear or reconnection instabilities.%
\footnote{
  Different nomenclature is used by different people.  Usually, these
  instabilities are studied in the magneto-hydrodynamic limit (MHD),
  which is the opposite limit of the collisionless plasmas studied
  in this paper, as MHD applies to physics on scales large compared to
  the (transport) mean free path.  However, there are analogous processes
  in the collisionless limit.  See, for instance, the analysis in
  Sec.\ 6.2.2 of Ref. \cite{galeev}.  When reading the plasma literature,
  however,
  it is important to keep in mind the difference between the MHD and
  collisionless limits since the physics can be different.
}

Consider the magnetic field configuration depicted in Fig.\ \ref{fig:tear}a.
By Ampere's Law, a stable, time-independent solution of this form
requires currents in the $x$ direction, also shown in
the figure.%
\footnote{
   The regions of current in such situations are referred to in the plasma
   literature as ``current sheets.''
}
Visualize the currents as being carried by wires.  Two parallel wires
with current in the same direction attract each other through
magnetic interaction.  There can then be an instability for the
wires to dimerize---that is, clump together, as shown in
Fig.\ \ref{fig:tear}b.  The magnetic fields then change correspondingly,
as indicated in the figure.
In the context of collisionless plasma theory, an analytic analysis
of this instability in a similar situation may be found in
Ref.\ \cite{galeev}.%
\footnote{
   Berger and Davidson \cite{berger&davidson}
   perform a stability analysis for a
   class of magnetic BGK solutions.  However, they only study stability
   within the subspace of $xy$ translation invariant configurations.
   As a result, their analysis would not find a tear/reconnection
   instability.  For the same reason, the tear/reconnection instability
   cannot appear in the original simulations of Davidson {\it et al.}\
   \cite{davidson&etal}, nor in
   more recent one-dimensional simulations
   by Yang, Arons, and Langdon \cite{yang2} of the Weibel
   instability in relativistic electromagnetic plasmas.
}
It has been observed in numerical simulations
of non-relativistic plasmas by Califano {\it et al.}\ \cite{califano},
which we will review in a moment.
The interesting qualitative feature of the tear/reconnection instability
is that it breaks the $xy$ translation invariance!  The non-perturbative
dynamics of fully grown plasma instabilities therefore has the
potential to isotropize the system on relatively short time scales.

\begin{figure}
\includegraphics[scale=0.40]{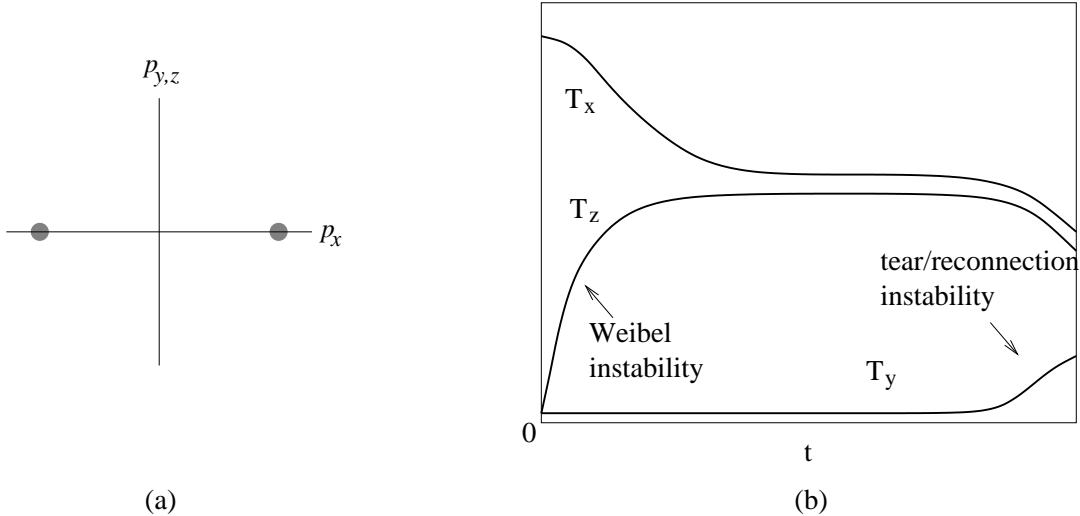}
\caption{%
    \label{fig:califano}
    (a) A two-stream initial condition for $f(\p)$.
    (b) A qualitative picture of the resulting evolution of the initial
    magnetic seed field (\ref{eq:califanoseed}).  The right half of
    figure (b) is a smoothed, qualitative sketch of the numerical results
    of Ref.\ \cite{califano}.  The left-hand side was not shown specifically
    in Ref.\ \cite{califano} but is pieced together from
    statements in the text of that reference and from earlier
    simulations of the Weibel instability in Ref.\ \cite{davidson&etal}.
}
\end{figure}

Califano {\it et al.}\ study the case of a hard particle
distribution corresponding to two similar, uniform, counter-streaming beams
of electrons which have a tiny thermal spread of velocities around the
beam velocities, as depicted in Fig.\ \ref{fig:califano}a.%
\footnote{
   Here and throughout, we have translated the coordinate labels
   used by Califano {\it et al}.\ \cite{califano}
   from $(x,y,z)$ to $(x,z,y)$, to make the dominant term in
   (\ref{eq:califanoseed}) depend only on $z$, as has been our
   convention in this paper.
   Their simulations are 2D+3V, meaning they treat 5 dimensions (2 space
   and 3 velocity) of 6-dimensional phase space.
   That is, they take configurations to be homogeneous
   in one spatial direction.
}
Additionally, rather than taking random initial conditions for the
seed magnetic fields, they take initial conditions of the form
\begin {equation}
   \B = \mbox{(small)}\times \sin(k_0 z) \, {\bm e}_y
        ~~+~~ \mbox{(much smaller random noise)} .
\label {eq:califanoseed}
\end {equation}
The first term seeds a single unstable mode that is $xy$ translation
invariant and grows into something like the non-linear magnetic BGK mode
discussed earlier.  The second term is even smaller, and seeds the
subsequent tear/reconnection instability of the magnetic BGK mode.
They made the two terms different sizes because they wanted to clearly
see a magnetic BGK mode and its subsequent demise, rather than having
it all jumbled together.  A qualitative depiction of the subsequent
evolution, pieced together from Refs.\ \cite{davidson&etal,califano}, is
shown in Fig.\ \ref{fig:califano}b.  $T_x$, $T_y$, and $T_z$ are
measures of the total kinetic energies along the $x$, $y$, and $z$
axis respectively, defined in terms of $m v_x^2$, $m v_y^2$, and
$m v_z^2$.  Initially, almost all of the electron kinetic energy is
in the $x$ direction.  Due to the Weibel instability, the first term
of (\ref{eq:califanoseed}) then seeds the growth of a single, dominant
unstable mode.  The $\B$ field of that mode points in the $y$
direction and so can deflect the particles in the $xz$ plane but does
not affect $v_y$.  The growth of this instability eventually saturates
in a nonlinear magnetic BGK-like state.  Then, at late times, the
second term of (\ref{eq:califanoseed}) grows due to the tear/reconnection
instability.  This instability drives the system toward isotropization of
$T_x$, $T_y$, and $T_z$.  Unfortunately, Califano {\it et al.}'s
simulations end before one can see for sure whether there is complete
isotropization.

One may wonder whether plasma instabilities always isotropize a
collisionless plasma.  Refs.\ \cite{honda,taguchi,semtoku,lee&lampe}
numerically study parity-asymmetric initial conditions which do not
isotropize.  In particular, Honda {\it et al.}\ \cite{honda} study
the initial condition of a homogeneous beam of fast electrons in
a plasma of slow electrons (and slower ions).
As time evolves, they find that the Weibel instability causes clumping of
the beam in 
the transverse direction.  The resulting current filaments then
attract each other magnetically (again just as two parallel wires carrying
current in the same direction), and the filaments begin to merge,
creating larger and larger filaments.  Finally, there is only one
filament left in their simulation volume, and the plasma stabilizes in this
configuration.  This final configuration is clearly anisotropic and
corresponds to an equilibrium
solution to the Vlasov equations originally found by Bennett in 1934
\cite{bennett},
which we review for the ultra-relativistic case (where it is simpler)
in Appendix \ref{app:bennett}.

Because of the parity-noninvariance of
the initial state,
the simulations of Honda {\it et al.} could end up with a
parity-noninvariant final state, described by the Bennett self-pinching
filament.  It is possible, in contrast, that generic
parity-{\it invariant} initial
conditions may lead to isotropization through collisionless processes,
as suggested by the results of Califano {\it et al.}


\section {Lower bound for the complete thermalization time}
\label{sec:bound}

In the introduction, we pointed out the embarrassing fact that theory has so
far been unable to determine even the power $n$ in the parametric relation
$\tau_{\rm eq} \sim \alpha^{-n} \Qs^{-1}$ for weak coupling, where $\tau_{\rm
eq}$ is the local thermalization time.  In this section, we point out that
there is a simple lower bound on $n$.  Consider the last part of
thermalization, where the system is approximately but not fully equilibrated.
At this point, we can use near-equilibrium results for the time scales
associated with approach to equilibrium.  Roughly, the characteristic time
scale is then simply the equilibrium time scale for (i) large-angle
deflections of a given particle's direction, and (ii) number-changing
processes such as hard gluon bremsstrahlung and $q\bar q$ creation or
annihilation.  Both
of these processes have been considered in detail in the literature
analyzing
near-equilibrium transport processes, such as the calculation of the
quark-gluon plasma shear viscosity \cite{shear,shear1,shear2}.
Parametrically, the time scale
for both these types of processes is
\begin {equation}
   \bar t \sim \frac{1}{\alphas^2 T}
\end {equation}
up to a factor of $\ln(1/\alphas)$, which we will not bother to keep track
of.  Full local equilibration cannot happen any faster than this, and
so
\begin {equation}
  \tau_{\rm eq} \gtrsim \frac{1}{\alphas^2 T} \,.
\label {eq:tau1}
\end {equation}
If the system has fully equilibrated at time $\tau_{\rm eq}$
during the
1-dimensional expansion, then the energy density at that time,
\begin {equation}
   {\cal E}(\tau_{\rm eq}) \sim \frac{{\cal E}(0)}{\Qs \tau_{\rm eq}}
                           \sim \frac{\Qs^4}{\alphas \Qs \tau_{\rm eq}}
   \,,
\end {equation}
must be $T^4$, so that
\begin {equation}
   T \sim \frac{\Qs}{(\alphas \Qs \tau_{\rm eq})^{1/4}} \,.
\label {eq:tau2}
\end {equation}
Combining (\ref{eq:tau1}) and (\ref{eq:tau2}), we obtain the
lower bound
\begin {equation}
   \Qs \tau_{\rm eq} \gtrsim \alphas^{-7/3} .
\end {equation}

Note that the result $\Qs \tau_{\rm eq} \sim \alphas^{-13/5}$ of the
original bottom-up scenario \cite{BMSS} satisfies this bound.  In that
case, thermalization happened slower than the bound because the energy
of the system got hung up for a time in hard modes,
before it could equilibrate by cascading down into softer modes.
The effects of plasma instabilities might speed up
equilibration, and it is conceivable that the real answer could saturate
the lower bound.  Even if it does, one would still like to know all the
various time scales associated with the different stages of equilibration,
and what happens at each stage.
The time scale for complete local equilibration might not be the most
relevant time scale for understanding the physics of heavy ion collisions.


\section{Conclusions}
\label{sec:conclusion}

In this paper, we have examined the effect of non-Abelian self-interactions
on growing plasma instabilities in an anisotropic non-Abelian plasma.
By looking at the effective potential for $z$-dependent magnetic fields
in the presence of an anisotropic distribution $f(\p)$ of hard particles,
we have given suggestive arguments that non-Abelian interactions
drive the growing instabilities to become Abelian once they grow
large enough.  Numerical simulations of a toy model 1+1 dimensional
gauge theory also show this behavior.  We conjecture that (i) the
plasma instabilities of SU(2) gauge theory grow into those of
traditional (relativistic) U(1) gauge theory, and (ii) the plasma
instabilities of QCD grow into those of U(1)$\times$U(1) gauge theory,
which is like traditional plasma theory but with two copies of
electromagnetism.  In the introduction, we asked what sets the scale for
how large plasma instabilities grow.  Our proposed answer is the Abelian
one: the scale $A \sim p_{\rm soft}/g$ for hard particles to be affected
non-perturbatively, rather than the scale $A \sim k_{\rm soft}/g$
for soft modes to have significant non-Abelian interactions.

Clearly, these conclusions need to be verified in 3+1 dimensional models,
and with the full HTL effective action (\ref{eq:Seff1}).  In order to
better understand the development of plasma instabilities once they
are grown, full simulations of the full non-linear Vlasov equations
are needed.  Further simulations of traditional U(1) theories would
provide useful information.  Simulations of U(1)$\times$U(1) theories
would be interesting as well.
Of course, full simulations of non-linear SU(3) Vlasov
equations would be even better.
It is also possible that non-equilibrium QCD plasma physics
could be studied in simulations
of pure classical gauge theories on the lattice, in situations where
there is a momentum scale separation so that excitation of
some modes can be thought of
as hard ``particles'' and others as soft fields.


\begin{acknowledgments}

We would like to thank Larry Yaffe, Guy Moore, Mike Strickland,
Francesco Califano, and Jonathan Arons for useful conversations.
This work was supported, in part,
by the U.S. Department of Energy under Grant No.~DE-FG02-97ER41027.

\end{acknowledgments}

\appendix


\section{The effective potential for \boldmath$\A=\A(z)$}

\subsection{For $\A = \A(z)$}
\label {app:Veff}

In this appendix, we will derive the result (\ref{eq:Vz}) for
the effective potential when $\A = \A(z)$.  We will begin,
however, with a more general discussion.
We find it advantageous to use a slightly different form of
the effective action (\ref{eq:Seff1}):
\begin {subequations}
\label {eq:Seff}
\begin {equation}
   S_{\rm eff} =
   - \int_x \fourth {F^a}_{\mu\nu} F^{a\mu\nu}
   - c g^2 \int_x \int_\p
     \frac{f}{p} \, {W^a}_\alpha W^{a\alpha} ,
\end {equation}
where
\begin {equation}
   W_\alpha = W_\alpha(x,\v)
   \equiv \frac{v^\mu}{v\cdot D} \, F_{\mu\alpha}(x) .
\end {equation}
\end {subequations}
This form can be obtained from (\ref{eq:Seff1}) by using the
anti-symmetry of the operator $\v\cdot\D$ in $x$/color
space to move one factor of $(\v\cdot\D)^{-1}$ from one $F$ to the
other.%
\footnote{
   In more detail, the adjoint-representation operator $\D$ is
   a real anti-symmetric operator in $x$/color space.  That means
   that the inverse $(v\cdot D)^{-1}$ of $v \cdot D$ is anti-symmetric
   as well.
   Now think of operating on some state $|s\rangle$ in configuration/color
   space, which represents a real function $s^a(x)$ with a single adjoint
   color index.
   Then the anti-symmetry can be used to rewrite
   $\langle s' | (\v\cdot\D)^{-2} | s \rangle =
   - \left[ (\v\cdot\D)^{-1} |s'\rangle \right]^\top
            (\v\cdot\D)^{-1} |s\rangle$.
   Taking $s'$ and $s$ of the form $v^\mu {F^a}_{\alpha\mu}$
   and $v^\nu {{F^b}_\nu}^{\alpha} = - v^\nu {F^{b\alpha}}_\nu$, one can
   then obtain (\ref{eq:Seff}) from (\ref{eq:Seff1}).
}
Here and throughout this appendix, we will not bother keeping
track of the $\epsilon$ prescriptions for retarded behavior.
We note in passing that, in the isotropic case, Iancu \cite{iancuH}
has discussed a
Hamiltonian formulation which is similar to (\ref{eq:Seff}).

Evaluating this action for static
configurations $\A(\x)$ in $A_0=0$ gauge, we obtain
\begin {subequations}
\label {eq:Veff}
\begin {equation}
   V_{\rm eff} =
   \int_\x \fourth F^a_{ij} F^a_{ij} + {\Delta V} ,
\end {equation}
where
\begin {equation}
   \Delta V \equiv
     c g^2 \int_\x \int_\p
     \frac{f}{p}\, W^a_k W^a_k ,
\label {eq:dVeff}
\end {equation}
and now
\begin {equation}
   W_k = W_k(\x,\v)
   = \frac{v_i}{\v\cdot\D} \, F_{ik}(\x) .
\end {equation}
\end {subequations}

Now we specialize to $\A = \A(z)$.
One can then make use of an identity noted by
Blaizot and Iancu \cite{BIwaves}, which is that
\begin {equation}
   v\cdot D \left(A^\alpha - \frac{n^\alpha v\cdot A}{n\cdot v}\right)
   = v^\nu {F_\nu}^\alpha
\label {eq:BI}
\end {equation}
if $A = A(n\cdot x)$ is only a function of $n\cdot x$, for some
constant 4-vector $n$.  This identity can be checked by explicitly
expanding all the terms on both sides.
Now apply $(v\cdot D)^{-1}$ to both sides of (\ref{eq:BI})
to get
\begin {equation}
   W^\alpha =
   A^\alpha - \frac{n^\alpha v\cdot A}{n\cdot v} .
\label {eq:copout}
\end {equation}
In our case of $\A = \A(z)$ with $A_0 = 0$, the identity is
\begin {equation}
   W_k = A_k - \delta_{kz} \, \frac{\v\cdot\A}{v_z} .
\end {equation}
Substituting into (\ref{eq:dVeff}), we then get that $\Delta V$
is quadratic in $\A(z)$.  But that means that it must be the same
as its expansion to quadratic order in $\A$, and so must be the same
as the result in the linearized theory, giving
\begin {equation}
   \Delta V{\bm[}\A(z){\bm]}
   = \int_\x \half \, A^a_i \Pi_{ij} A^a_j .
\end {equation}
In $\k$ space, one may therefore write
\begin {equation}
   \Delta V{\bm[}\A(z){\bm]}
   = \int_\k \half \, A^a_i(\k)^* \, \Pi_{ij}(0,\hat\k) \, A^a_j(\k) .
\end {equation}
Since the Fourier transform of $\A = \A(z)$ has support only for
$\k$'s proportional to $\hat\e_z$, we can replace $\Pi^{ij}(0,\hat\k)$
by the matrix of constants $\Pi_{ij}(0,\hat\e_z)$.
The effective potential for $\A = \A(z)$ is then local
in $\x$ and may be written in the form (\ref{eq:Vz}).


\subsection{For SU(2) with $k_\perp=0$, $k_z\to 0$ and $A_z=0$}
\label{app:topo}

In this section, we will simplify the potential ${\cal V}$ of
(\ref{eq:pot}) under the assumptions that 
(i) $\A$ lies in an SU(2) subgroup of color SU(3)
and (ii) $A_z = 0$.
First, let's consider the restriction
to an SU(2) color subgroup.  Then $f^{abc}$ can be replaced
by $\epsilon^{abc}$, giving
\begin {eqnarray}
   {\cal V}_{\rm SU(2)} &=&
       \fourth \, g^2 \Bigl[
        (A^b_i A^b_i)^2 - A^b_i A^c_j A^c_i A^b_j
       \Bigl]
       - \half \, \mu^2 (A^a_x A^a_x + A^a_y A^a_y)
\nonumber\\
   &=& \fourth \, g^2 \Bigl\{ [\tr(\Amat^\top\Amat)]^2
                              - \tr[(\Amat^\top\Amat)^2] \Bigl\}
       - \half \, \mu^2 \tr(\Amat^\top P^{(xy)} \Amat) ,
\label {eq:pot2}
\end {eqnarray}
where $\Amat$ is the $3\times3$ matrix of $A_{ia}$,
\begin {equation}
   {\cal A} = \begin{pmatrix}
                 A_x^1 & A_x^2 & A_x^3 \\[2pt]
                 A_y^1 & A_y^2 & A_y^3 \\[2pt]
                 A_z^1 & A_z^2 & A_z^3
               \end{pmatrix} .
\end {equation}
$P^{(xy)}$
is the projection operator onto the $xy$ subspace of directions $i$,
\begin {equation}
   P^{(xy)} = \begin{pmatrix} 1 & & \\ & 1 & \\ & & 0 \end{pmatrix} .
\end {equation}
The potential is symmetric under (i) spatial rotations in the $xy$ plane,
and (ii) color rotations.  These symmetries can be summarized as
\begin {equation}
   {\cal A} \to R {\cal A} C^\top ,
\end {equation}
where $C$ is a color rotation for the adjoint representation of SU(2),
represented by any $3\times3$ real, orthogonal matrix with $\det C = 1$, and
$R$ is a $3\times3$ real, orthogonal matrix representing a rotation in
the $xy$ plane.
By a color transformation, we can assume
without loss of generality that the matrix $\Amat$ is symmetric.%
\footnote{
  Using singular value decomposition, we can write
  $\Amat = L^\top D R$ for any $\Amat$, where $L$ and $R$ are
  orthogonal matrices and $D$ is a diagonal matrix.  Then
  $\Amat R^\top L = L^\top D L$ is symmetric.
  So, a color rotation
  of $\Amat$ by $R^\top L$ makes $\Amat$ symmetric.
}
Now make the remaining simplifying restriction that
$A_z = 0$.  Then the symmetric ${\cal A}_{ia}$ is zero except for
$i = x,y$ and $a = 1,2$.  We can then diagonalize the symmetric ${\cal A}$
in this $2\times2$ subspace by a simultaneous space/color rotation of the
form ${\cal A} \to R {\cal A} R^\top$.  So, without further loss of
generality, we may write
\begin {equation}
   {\cal A} = \begin {pmatrix} \phi_1 & & \\ & \phi_2 & \\ & & 0 \end{pmatrix}
   .
\label{eq:Aphi}
\end {equation}
This gives (\ref{eq:Aform}).


\section{Unstable, non-Abelian, static waves}
\label {app:wave}

In this appendix, we show the existence of non-linear, non-Abelian,
static wave solutions
to the effective theory (\ref{eq:Seff}) of the soft modes.
These solutions
are classically unstable, however, to the Abelianization discussed in
this paper.  In the zero momentum limit ($k\to 0$), we will see that they
correspond to the saddle points in Fig.\ \ref{fig:V}.

The solutions are simple generalizations of propagating non-linear
wave solutions found for equilibrium plasmas by Blaizot and Iancu
\cite{BIwaves}.  We will look for solutions of the form
$A = A(K\cdot x)$, for some 4-vector $K$.  [We will shortly specialize to
$K$ in the $z$ direction, so that $A = A(z)$.]  As discussed in Appendix A,
Blaizot and Iancu then showed that, in this case,
the only effect of hard particles
at leading order is to induce the HTL self-energy for the soft
fields.  The soft equation of motion is then
\begin {equation}
   D_\nu F^{\mu\nu} = - \Pi^{\mu\nu} A_\nu .
\label {eq:BIeq1}
\end {equation}
Following one of the possibilities studied by
Blaizot and Iancu, let's focus on an SU(2) subgroup of the gauge
group and further
restrict attention to gauge fields of the form
\begin {equation}
   A^\mu_a(n\cdot x)
   = \bigl[ \epsilon^\mu_{(1)} \delta_{a1} + \epsilon^\mu_{(2)} \delta_{a2}
        \bigr] h(K\cdot x) ,
\label {eq:hansatz}
\end {equation}
where $\epsilon^\mu_{(1,2)}$ is a basis of spatial polarizations orthogonal
to $\k$.
[This is essentially just (\ref{eq:Aphi}) with $\phi_1 = \phi_2 = h(z)$,
if one chooses $K$ along the $z$ axis.]
The equation of motion (\ref{eq:BIeq1}) then becomes
\begin {equation}
  (\omega^2-k^2) h'' + \Pi_T(\omega,\p) \, h + g^2 h^3 = 0 ,
\label {eq:BIeq2}
\end {equation}
where $h''$ indicates the second derivative of $h$ with respect
to its argument $K\cdot x$.

Blaizot and Iancu considered eq.\ (\ref{eq:BIeq2}) for $\Pi_T$ positive,
which is the situation for propagating modes in equilibrium
situations.  There are no non-trivial
static solutions ($\omega = 0$) in this case because
$\Pi_T(0,\k) = 0$ in equilibrium.
The difference in this paper is that we are interested in cases
where $\Pi_T(0,\k)$ can be negative because of the Weibel instability
for anisotropic distributions.  As a result, static solutions can
exist to (\ref{eq:BIeq2}).  Taking $\k$ to be in the $z$ direction,
and writing $\Pi_T(0,\k) = -\mu^2 < 0$, the static ($\omega=0$) case
of (\ref{eq:BIeq2}) is
\begin {equation}
  - \frac{d^2h}{d z^2} - \mu^2 h + g^2 h^3 = 0 .
\label {eq:heq}
\end {equation}
The solutions are
\begin {equation}
  h(z) =
  \frac{\mu}{g} \left(\frac{2m}{1+m}\right)^{1/2}
  \operatorname{sn}\left(\frac{\mu z}{(1+m)^{1/2}} \,\Big|\, m\right) ,
\label{eq:hsoln}
\end {equation}
for any $m$ with $0 < m < 1$.
Here $\operatorname{sn}(u|m)$ is the Jacobi elliptic sine of
argument $u$ and parameter $m$.

Now note that (\ref{eq:heq}) is just the equation for static solutions
of a scalar theory with potential
$V(h) = - \half \mu^2 h^2 + \fourth g^2 h^4$.
This double-well potential is simply the potential of Fig.\ \ref{fig:V}
along the diagonal $\phi_1 = \phi_2$.  The minima of $V(h)$ correspond
to the saddle points of $V(\phi_1,\phi_2)$.  In the ansatz
(\ref{eq:hansatz}), corresponding to $\phi_1=\phi_2$, we simply
cannot see the unstable directions of the saddle point.

For $m \ll 1$ in (\ref{eq:hsoln}), the solution has small-amplitude
variations around $h=0$ as one changes $z$.  As $m \to 1$, the
solution becomes a regular array of far-separated kinks and anti-kinks,
where the standard kink solution, centered on the origin, is
$h(z) = (\mu/g) \, \tanh(\mu z/\sqrt2)$.  In this case, the kinks interpolate
back and forth diagonally between the saddle points of Fig.\ \ref{fig:V},
and in most of space, the field is nearly constant near one saddle point
or the other.  And so the constant-field instabilities discussed in the
main text will locally afflict these solutions.  For smaller $m$,
where the kinks and anti-kinks begin to overlap, the potential
energy can be further minimized by annihilating pairs of kinks and anti-kinks
(that is, by dimerization): so again, these represent unstable solutions.


\section{Can soft excitations stop instability growth?}
\label {app:qualitative}

In this appendix, we give some qualitative arguments against the concerns
raised in section \ref{sec:hangup}.  We will argue that one should not
expect effective masses generated by the soft sector to be able to
prevent the growth of the system down along the valleys of the
effective potential shown in Fig.\ \ref{fig:V}.  Because we discuss
theories in different dimensions in this paper, we shall keep the
spatial dimension $d$ general.  For simplicity, we will restrict our
discussion to the purely scalar theory of our warm-up toy model
(\ref{eq:ScalarToy}).

\begin{figure}
\includegraphics[scale=0.50]{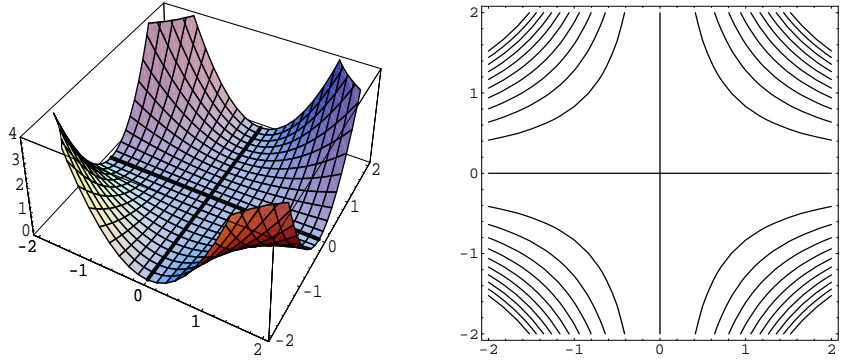}
\caption{%
    \label{fig:Vhung}
    Like Fig.\ \ref{fig:V} but including a hypothetical contribution from
    soft modes that cancels the $-\mu^2 A^2$ due to hard particles.
    This potential is zero along the axis.  Qualitatively, the creation of
    just slightly more power in soft modes would cause it to curve up
    instead.
}
\end{figure}

Imagine that the system starts with very tiny random magnetic noise,
and the modes with $k \le \mu$ then grow due to the instability.
In a time of order $\mu^{-1}$ (neglecting factors of logarithms),
the instabilities grow to the size
of the soft non-perturbative scale $\mu/g$.
Let's start by supposing that the growth of the system
{\it did}\/ hang up at this scale.
That is, suppose that the excitation of soft modes has built up to
a point such that their contribution to the effective potential now
cancels the destabilizing $-\mu^2$ quadratic coefficient
contributed by the hard modes, so that, crudely speaking,
the {\it total}\/
effective potential at this
time looks somewhat like Fig.\ \ref{fig:Vhung} rather than
Fig.\ \ref{fig:V} (which accounted only for the effects of the hard
particles).
The potential energy density released at this point would
be of order $\mu^4/g^2$, which will also be the kinetic energy density of
the soft field.  At least initially, this energy would be
in Fourier components with $k \sim \mu$ and with non-perturbatively large
occupation numbers of order $1/g^2$.  If the system were hung up,
however, then it would quickly start to equilibrate by the strong
non-perturbative self-interactions of these modes.

If the hung-up system were equilibrated, then the kinetic energy density
$\mu^4/g^2$ would be $T^{d+1}$, giving $T \sim (\mu^{4}/g^{2})^{1/(d+1)}$.
This gives $T \gg \mu$ for weakly-coupled theories $\mu^{d-3} g^2 \ll 1$
in $d$ spatial dimensions.
The energy would then be dominated by modes with momentum of order
$T \gg \mu$.  The drive toward thermalization would therefore
push the energy
of the system from modes with $k \sim \mu$ to modes with higher and higher
momenta, until the system eventually thermalized (if it in fact remained
hung up at $\phi \sim \mu/g$).

Let $\Lambda$ be the momentum scale that dominates the total kinetic
energy at any time, which starts at order $\mu$.
Due to the drive of the system toward thermalization,
$\Lambda$ would grow several times in a time of order $\mu^{-1}$.
Now consider a time when $\Lambda \gg \mu$.  At this time, the
kinetic energy density $\mu^4/g^2$, which must be of order $k^2 \phi^2$,
is distributed among modes with momentum $k\sim\Lambda$.  So now
$\phi \sim \mu^2/g\Lambda$.  By mean field theory, the corresponding
contribution to the squared effective mass is
\begin {equation}
   (\Delta m^2)_\Lambda \sim g^2 \phi^2
   \sim \frac{\mu^4}{\Lambda^2} .
\label {eq:msoft}
\end {equation}
Were $\Lambda \sim \mu$, this might or might not be enough to compensate
for the unstable $-\mu^2$ in the potential.  However, since thermalization
drives $\Lambda$ larger and larger, one will soon have
$-\mu^2 + (\Delta m^2)_\Lambda < 0$, and so instability to further growth,
in a time scale of order
$\mu^{-1}$.

There was an implicit assumption above that the momentum scale
$\Lambda$ which dominates the kinetic energy is also the scale
giving the dominant contribution to $\Delta m^2$.
In low dimensions, this need not be the case.
But even if all the energy were in modes $k \sim \mu$,
corresponding to $\Lambda \sim \mu$ above,
we have seen that the
resulting $(\Delta m^2)_\mu$ is order $\mu^2$.
Once most of the energy has left modes $\mu\sim k$, then
$(\Delta m^2)_\mu$ will be $\ll \mu^2$.  So neither
$(\Delta m^2)_\mu$ nor $(\Delta m^2)_\Lambda$ will
be enough to stop the continued growth of the instability.%
\footnote{
  Here's a different form of the analysis.  Consider a system that has
  equilibrated for modes $k \lesssim \Lambda \ll T$, so that the occupation
  numbers are $f(k) \sim T/E_k$ for $k \lesssim \Lambda$ and are small
  for larger $k$.  Then the kinetic energy density is order
  $\int d^dk \> E_k f(k) \sim T \Lambda^{d}$, which we set to
  $\mu^4/g^2$ as above.  From this, we obtain $T \sim \mu^4/g^2 \Lambda^d$.
  The contribution to the effective mass from all the soft modes is
  $\Delta m^2 \sim g^2 \int d^dk \> f(k)/E_k \sim g^2 T \int d^d k / E_k^2$.
  Using the previous expression for $T$, this is
  $\Delta m^2 \sim (\mu^4/\Lambda^d) \int d^d k / E_k^2$.
  For $d>2$, the UV contribution dominates, giving
  $\Delta m^2 \sim \mu^4/\Lambda^2$.
  For $d < 2$, the IR contribution dominates, giving
  $\Delta m^2 \sim \mu^{d+2}/\Lambda^d$.  For $d=2$,
  $\Delta m^2 \sim (\mu^4/\Lambda^2) \log(\Lambda/\mu)$.
  In all of these cases, $\Delta m^2 \ll \mu^2$ when $\Lambda \gg \mu$.
}


\section{Lattice version of toy model}
\label {app:lattice}

The lattice version of our 1+1 dimensional toy model,
with spatial lattice spacing $a$,
consists of the following elements.  There are adjoint scalar
fields $\phi_x$ and $\phi_y$, corresponding to $A_x$ and $A_y$,
which live on the sites of the spatial lattice, represented as
$N{\times}N$ traceless Hermitian matrices of su($N$).
There are  SU($N$) group elements $U$ living on the spatial links,
corresponding to $\exp(i a  A_z)$ and represented as
$N{\times}N$ unitary matrices.
Then there are the conjugate momenta $\Pi_x$ and $\Pi_y$
(corresponding to $\dot A_x = -E_x$ and $\dot A_y = -E_y$)
and $E$ (corresponding to $E_z$), which are all represented as
$N{\times}N$ traceless Hermitian matrices living on the temporal links.
In this appendix, we take $g=1$ and $\mu=1$.
For time step $\eps$, our evolution equations are
\begin {eqnarray}
   \phi_{\alpha,s}(t+\eps)
   &=& \phi_{\alpha,s}(t) + \eps \, \Pi_{\alpha,s}(t+\half\eps) ,
\\
   U_{s+\half}(t+\eps) &=& U_{s+\half}(t) \>
     \exp\Bigl(-i \eps a \, E_s(t+\half\eps)\Bigr) ,
\\
   \Pi_{\alpha,s}(t+\half\eps) &=&
   \Pi_{\alpha,s}(t-\half\eps)
   + \eps \biggl(
       a^{-2} \Bigl(
            U_{s+\half}^\dagger \phi_{\alpha,s+1} U_{s+\half}
            - 2 \phi_{\alpha,s}
            + U_{s-\half} \phi_{\alpha,s-1} U_{s-\half}^\dagger
       \Bigr)
\nonumber \\ && \hspace{9em}
       + \mu^2 \phi_{\alpha,s}
       - [\phi_{\bar\alpha,s},[\phi_{\bar\alpha,s},\phi_{\alpha,s}]]
     \biggl)_{t} ,
\\
   E_s(t+\half\eps) &=&
   E_s(t-\half\eps)
   - \frac{\eps}{a} \sum_{\alpha}
      i \big[ \phi_{\alpha,s} \> , \> 
          U_{s+\half}^\dagger \phi_{\alpha,s+1} U_{s+\half} \big]_t ,
\end {eqnarray}
where $s$ is an integer site index,
$\alpha$ runs over $x$ and $y$,
and $\bar\alpha$ denotes $y$ and $x$, respectively.
Gauss's Law is
\begin {equation}
   a^{-1} \Bigl(
      E_s(t-\half\eps)
      - U_{s-\half}(t) \, E_{s-1}(t-\half\eps) \, U^\dagger_{s-\half}(t)
   \Bigr)
   + \sum_\alpha i \big[ \phi_{\alpha,s}(t) \, , \, 
                         \Pi_{\alpha,s}(t-\half\eps) \big]
   = 0 .
\end {equation}

The approximately conserved energy (which is exactly conserved
for $\eps\to 0$) is
\begin {multline}
   {\cal E} = a \sum_s \biggl[
     \tr(E_s^2)
     + \sum_\alpha \tr(\Pi_{\alpha,s}^2)
     + \frac1{a^2} \sum_\alpha
         \tr \left( \Bigl(\phi_{\alpha,s} 
            - U_{s-\half} \phi_{\alpha,s-1} U_{s-\half}^\dagger
         \Bigr)^2 \right)
\\
     - \mu^2 \tr(\phi_{x,s}^2+\phi_{y,s}^2)
     + \tr\Bigl( \bigl(i [\phi_{x,s},\phi_{y,s}]\bigr)^2 \Bigr) 
   \biggr].
\label {eq:energy}
\end {multline}
In our simulations, the total energy of the system is very close to zero,
because we start with very small initial fields.  One can get a measure of
how well energy is conserved in our simulations by comparing the
change ${\cal E}(t)-{\cal E}(0)$ in the energy with time to the
size of the potential energy $V(t)$ given by the last three terms of
(\ref{eq:energy}).  By this measure, the
simulations presented in this paper conserve energy at the level of
0.1\%.  We have checked that reducing the time step $\epsilon$ from
the value $\epsilon=0.001$ used in our simulations
(i) does not significantly change any of our
simulation results, and (ii) indeed further improves energy conservation.

Our initial conditions (\ref{eq:initial}) are
\begin {equation}
   U_{s+\half}(0) = 1,
   \qquad
   \phi_{\alpha,s}(0) = 0 ,
   \qquad
   E_s(-\half\eps) = 0 ,
\end {equation}
\begin {equation}
   \bigl\langle \Pi^a_{\alpha,s}(-\half\eps) \,
                \Pi^b_{\beta,s'}(-\half\eps)
   \bigr\rangle
         = \frac{\Delta^2}{a} \,
           \delta^{ab} \delta_{\alpha\beta} \delta_{ss'} ,
\label {eq:whiteL}
\end {equation}
where $\langle \xi \xi \rangle = \Delta^2/a$ means choose $\xi$ to be
a Gaussian random number centered on zero with standard deviation
$\sigma = \Delta/a^{1/2}$.
In our simulations, $\sigma = 0.01$.
The initial energy density $\rho_0$ is of order
\begin {equation}
   \rho_0 \sim \dot\phi^2 \sim \sigma^2
\label {eq:rho0}
\end {equation}
and is dominated by stable UV modes.

We should check that our simulations will not run into a common
difficulty in naively applying classical evolution equations to many
problems, which is exemplified by the ultraviolet catastrophe of the classical
blackbody spectrum.  In thermal equilibrium, classical field
theory (unlike its quantum counterpart), populates arbitrarily high
momentum modes with real excitations.  The interaction of these
excitations with low-momentum modes of interest can cause arbitrarily
large effects, in the continuum limit, on effective masses, damping
rates, and other properties of the soft fields. (See, for example,
Refs.\ \cite{alpha5,hotB}.)
Our simple white noise spectrum (\ref{eq:whiteL}) populates arbitrarily
high-momentum modes, so one might worry that there could be a similar
problem here.%
\footnote{
  If there were, it could be fixed, of course,
  by cutting off the spectrum of the
  initial fluctuations.
}
As an example, let's check the effective masses of soft modes due to
interacting with UV modes of our lattice.  Such masses arise from
diagrams like Fig.\ \ref{fig:loop}.  There are a number of ways
to compute their scale.  Treating the UV lattice modes with kinetic theory,
for example,
\begin {equation}
   \Delta m^2 \sim g^2 \int_\p \frac{f_\p}{p} \,.
\end {equation}
where the integration is restricted to the UV modes whose effects are of
interest.
Previously in this paper, $f_\p$ has referred to hard particles.
In this context, it refers instead to the UV modes of our classical
continuum effective field theory of the ``soft'' modes.  The energy
density of these UV modes is $\rho \sim \int_\p E_p f_\p$.
Parametrically, then,
\begin {equation}
   \Delta m^2 \sim \frac{g^2 \rho}{p^2} \sim g^2 \rho a^2 .
\end {equation}
The effects on the soft physics of interest will be small if
$\Delta m^2 \ll \mu^2$.  Adopting our conventions $g=1$ and $\mu=1$,
and using (\ref{eq:rho0}),
this condition becomes
\begin {equation}
    \sigma^2 a^2 \ll 1 .
\end {equation}
In our simulations, $\sigma^2 a^2 = 2.5 \times 10^{-7}$.


\section{Ultra-relativistic Bennett self-pinch}
\label {app:bennett}

Traditional plasma physics is often complicated by the presence of multiple
scales arising from the fact that ions are much heavier than electrons.  Such
complications vanish for electron-positron plasmas, or quark-gluon plasmas.
In this appendix, we will summarize the self-pinching solutions to the
collisionless Vlasov equations found by Bennett \cite{bennett} in the
simplifying situation where opposite charges have the same mass.
This includes the ultra-relativistic limit, where particle masses are
ignored.
For simplicity of notation, we will focus on Abelian gauge theory.
A class of self-pinching solutions for a plasma of charges $\pm e$
and mass $m$ is given by%
\footnote{
  One may derive this equation by starting from the ansatz
  (\ref{eq:fbennett}) for some unknown function $\rho(r_\perp)$,
  then use Ampere's Law to determine ${\bm B}$, and then use
  the collisionless Boltzmann equation for $f_\pm$ to obtain a
  differential equation for $\rho$.  Solving that equation yields
  (\ref{eq:rhobennett}).  See Ref.\ \cite{bennett}.
}
\begin {subequations}
\label {eq:bennett}
\begin {equation}
   f_\pm(\x,\p) =
   \rho(r_\perp) \, F_\pm(\p)
\label {eq:fbennett}
\end {equation}
in cylindrical coordinates $(r_\perp,\phi,z)$,
where $F_\pm(\p)$ are boosted thermal distributions
\begin {equation}
   F_\pm(\p) =
   \frac{\exp\left[-\gamma_u(E_p \mp u p_z)/T\right]}{\gamma_u Z_m} ,
   \qquad
   Z_m = \frac{m^2 T}{2\pi^2} \, K_2\!\left(\frac{m}{T}\right)
\end {equation}
(with $Z_m \to Z_0 = T^3/\pi^2$
in the ultra-relativistic limit $T \gg m$),
and $\rho(r_\perp)$ is the radial profile
\begin {equation}
   \rho(r_\perp) =
   \frac{\rho_0}{(1+b\rho_0 r_\perp^2)^2} \,,
   \qquad
   b = \frac{e^2 \gamma_u u^2}{4T} \,.
\label {eq:rhobennett}
\end {equation}
\end {subequations}
The constants $\rho_0$, $T$, and $u$ are arbitrary.
One may check explicitly that
(\ref{eq:bennett}) satisfies the time-independent collisionless Vlasov
equations
\begin {equation}
   \v\cdot\grad f_\pm \pm e(\E + \v\times\B)\cdot\grad_\p f_\pm = 0 ,
\end {equation}
\begin {equation}
   \grad\cdot\E = e\int_\p (f_+ - f_-) ,
\end {equation}
\begin {equation}
   \grad\times\B = e \int_\p \v (f_+-f_-) ,
\end {equation}
with $\E=0$ and
\begin {equation}
  \B = \frac{4 T b \rho_0 r_\perp}{e \gamma_u u (1+b\rho_0 r_\perp^2)}
       \, {\bm e}_\phi .
\end {equation}
This solution represents a finite-width beam of positively charged
particles moving in the $+z$ direction and an exactly similar beam of
negatively charged particles moving in the $-z$ direction.  At their
centers, each of these beams has density $\rho_0$ and looks like
a thermal distribution with temperature $T$ boosted to a net beam
velocity of magnitude $u$.
The beam velocity $u$ can take any value $0<u<1$, even in the
ultra-relativistic limit, since $u$ is the {\em average}\/ velocity of
particles in the beam.


\end {document}